\documentclass[sigconf]{acmart}

\renewcommand\footnotetextcopyrightpermission[1]{} 
\AtBeginDocument{%
  \providecommand\BibTeX{{%
    \normalfont B\kern-0.5em{\scshape i\kern-0.25em b}\kern-0.8em\TeX}}}
    
\usepackage{tikz}
\usepackage{amsmath}
\usepackage{listings}
\usepackage{comment}
\usepackage{enumerate}
\usepackage{color}
\usepackage{relsize}
\usepackage{booktabs}
\usepackage{algorithm}
\usepackage{algpseudocode}
\usepackage{graphics}
\usepackage[toc,page]{appendix} 

\newcommand{\figspace}{\vspace{-0mm}}

\setcopyright{rightsretained}

\begin{document}

\title{BAASH: Enabling Blockchain-as-a-Service on High-Performance Computing Systems}

\author{Abdullah Al-Mamun$^\dagger$, Dongfang Zhao$^{\dagger \S}$}
\affiliation{%
  \institution{$^\dagger$University of Nevada, Reno \quad $^\S$University of California, Davis}
}


\begin{abstract}
The state-of-the-art approach to manage blockchains is to process blocks of transactions in a shared-nothing environment.
Although blockchains have the potential to provide various services for high-performance computing (HPC) systems,
HPC will not be able to embrace blockchains before the following two missing pieces become available:
(i) new consensus protocols being aware of the shared-storage architecture in HPC, and
(ii) new fault-tolerant mechanisms compensating for HPC's programming model---the message passing interface (MPI)---that is vulnerable for blockchain-like workloads.
To this end, we design a new set of consensus protocols crafted for the HPC platforms and a new fault-tolerance subsystem compensating for the failures caused by faulty MPI processes.
Built on top of the new protocols and fault-tolerance mechanism, 
a prototype system is implemented and evaluated with two million transactions on a 500-core HPC cluster, 
showing $6\times$, $12\times$, and $75\times$ higher throughput than Hyperldeger, Ethereum, and Parity, respectively.
\end{abstract}
\maketitle
\section{Introduction}

\subsection{Motivation}

Although cryptocurrency like Bitcoin~\cite{bitcoin} recently draws most of the public attention, in many cases leading the misconception of the equivalence between cryptocurrency and its underlying abstraction \textit{blockchain},
a blockchain is nothing but a unique symbiosis of various existing computer science techniques from distributed computing, database transactions, applied cryptography, and recently algorithmic game theory.
The simple yet elegant design of this abstraction,
therefore, brought about surging research interests from the above areas,
including several best paper awards in leading conferences.
Once again, we wanted to emphasize that blockchain is a system abstraction,
or more specifically, 
a distributed system with high availability and reliability and at the same time with low space utilization and performance on throughput,
which happens to be the infrastructure of popular cryptocurrencies like Bitcoin and Ethereum.
\textit{There is nothing that prevents us from modifying or optimizing blockchains for specific workloads, including but limited to scientific applications}.
For instance, although Bitcoin network maintains a full copy of data on all the participant nodes,
a hypothetical blockchain designed for scientific applications or high-performance computing (HPC) does not have to do so.
Admittedly, the preconception and popularity of Bitcoin are so strong that many people, understandably, call and, in fact, mean a cryptocurrency a blockchain or \textit{vice versa}.

The U.S. Department of Energy (DOE), to our knowledge, was one of the first federal agencies who realized the transformative potential of and actually funded, blockchain-related projects in HPC.
We do not list the funded projects here in this paper;
interested readers can easily find out those projects on the DOE website.
In the remainder of this subsection, we will elaborate on why blockchains,
although used to be thought of hype of digital coins,
in fact, represent a \textit{technical} novelty and have the potential to change many HPC aspects. 
Before that, we briefly review some key properties of blockchains.
In a blockchain-based distributed ledger, a piece of data (usually generated by a transaction) is persisted to the disk only after it is verified by the majority of the participating nodes in the network based on the system's consensus protocols.
Such a system removes the centralized components and builds strong trust over the linked hashing values in the entire history of the (transaction) data,
making the system extremely hard to compromise and achieving high resilience.

Several potential use cases~\cite{exascale-valduriez2015data} make blockchains a prominent candidate to be applied in future HPC infrastructures such as distributed caching, data provenance, and the fidelity of scientific discovery. 

\subsubsection{Distributed and persistent caching through blockchains}
Exascale systems are likely to have extreme power constraints leading to moving data anywhere, necessarily near to the processors, which is expensive but soluble by replicating or caching data distributedly through a distributed ledger. 
Besides, performance in exascale systems will require enormous concurrency, which will need data in each node to be in the same state through synchronization. 
Blockchain can enable this data synchronization with ultimate reliability in an autonomous fashion.
The Oak Ridge National Laboratory recently released a white paper~\cite{ornl_bc18} discussing a wide range of potential applications that can benefit from blockchains on the Oak Ridge Leadership Computing Facility.

\subsubsection{Reliable and traceable data provenance}
Another important scenario to realize the necessity of a blockchain-like distributed ledger in HPC systems is managing reliable data provenance. 
HPC systems (e.g., Cori~\cite{cori}) reply on data provenance to reproduce and verify the scientific experimental results generated during the application executions. 
Essentially, data provenance is a well-structured log for efficient data storage along with an easy-to-use query interface.
Data provenance is conventionally implemented through file systems~\cite{dzhao_cluster13,dzhao_tapp13} or relational databases~\cite{slee_icde17,spade_2012}.
Recent studies~\cite{aalmamun_bigdata18,xliang_ccgrid17} showed that blockchains could be leveraged to provide efficient and reliable provenance.

\subsubsection{Scientific data fidelity}

Data fidelity is of prominent importance for many scientific applications deployed to HPC systems,
as the data upon which scientific discovery rests must be trustworthy and retain its veracity at every point in the scientific workflow.
There have been more than enough incidents about data falsification and fabrication, causing the withdrawal of scientific publications. 
To this end, developing trustworthy data service for scientific computing and HPC systems
has been recently incentivized by various federal funding agencies,
such as the National Science Foundation~\cite{nsf_cici_2018} and the U.S. Department of Energy~\cite{doe_sbir_2017}.
A framework on leveraging blockchains to improve the fidelity of scientific applications and fidelity was proposed in~\cite{aalmamun_sc19}.

Given the aforementioned various HPC scenarios potentially involving blockchains,
we argue that having a general blockchain framework, or even better, a ready-to-use blockchain service, 
would not be a question of if but when.
The goal of this work is to share our early effort and results in a blockchain-as-a-service in HPC.
In the long run, we believe blockchains, just as parallel file systems, MPI, dockerized encapsulation, and recently machine-learning-as-a-service (MLaaS)~\cite{hqin_sc19}, would soon join the family of system tools and services in HPC.

\subsection{Challenges}

While blockchains have drawn much research interest in many areas such as 
cryptocurrency~\cite{crypto-algorand-Gilad:2017,crypto-bitcoin-ng-Eyal:2016,crypto-kogias2016} and smart government~\cite{SmartGov-olnes2016}, 
the high-performance computing (HPC) and scientific computing communities,
although regarding resilience as one of their top system design objectives, 
have not taken blockchains into their ecosystems due to various reasons such as the shared-storage system infrastructure of HPC systems and the MPI programming model for scientific applications.
All of the mainstream blockchain systems and frameworks assume the underlying systems are shared-nothing clusters with the TCP/IP network stack.  

In particular, although blockchain itself exhibits many research and application opportunities (comprehensive reviews available from~\cite{kzhang_icdcs18,msadoghi_ebdt18,blockbench_sigmod17}), one of the most impelling challenges for employing blockchains into HPC systems lies on the consensus protocols and serialized I/O subsystem.   
Existing consensus protocols used in mainstream blockchains are either based on intensive computation (the so-called proof-of-work, or POW) or intensive network communication (e.g., practical Byzantine fault tolerance, PBFT),
which are inappropriate for HPC and scientific computing in terms of both performance and cost. Besides, in the present solutions, the block processing is operated in a serialized manner among the peers in a network, 
which incurs unnecessary processing time and thus is inapplicable to a large-scale HPC cluster.  

\subsection{Proposed Solution}

In this paper, we present BAASH, a new blockchain framework, specially designed for HPC systems developed with parallel communication mechanisms through MPI~\cite{mpi}.
The goal of this new blockchain framework is to make the decentralized ledger fully compatible with the infrastructure of HPC systems by overcoming the shortcomings of the present consensus protocols through a specially crafted parallel communication layer built with MPI. We focus on enabling three challenging yet highly-desired features when designing such a \textit{\underline{b}lockchain-\underline{a}s-\underline{a}-\underline{s}ervice} framework for \underline{H}PC, namely BAASH,
which are elaborated as follows.

\subsubsection{In-memory \& shared-storage consensus protocol}

We design a consensus protocol that supports a parallel mechanism to achieve the highest possible throughput without compromising security. We aim to make all the participants fully trustworthy while not imposing costly computational overhead nor using extensive message passing among the nodes to achieve trust.   The consensus protocol is inspired by the traditional consensus protocols (i.e., PoW and PBFT) but with novel designs to meet the special needs in HPC systems. 
In BAASH, a block is double validated with two individual steps. \textit{First,} the block is validated with a custom proof-of-work (PoW) in each node. To minimize the block validation time, the PoW in BAASH consists of moderate difficulty (i.e., low nonce) because, in the HPC cluster, all the nodes are at least partially trustworthy and will be verified before getting added to the network. \textit{Second}, as low nonce is used in PoW at the first step, further measures are considered to remove the notion of low security of PoW. To close any security gap further, we leverage the idea of the traditional PBFT with essential modification to facilitate parallel consensus achieving process through MPI---the de facto communication middleware in HPC and many scientific applications. 
If the majority of the compute nodes (i.e., at least 51\%) are unable to reach consensus about the validity of a block, the remote storage then participates in the block validation process.

\subsubsection{Parallel block processing}

To further improve the performance of the new consensus protocol, we propose a parallel approach for processing the data blocks through MPI (i.e., the mpi4py library~\cite{mpi4py}) to prepare the system prototype compatible with the HPC infrastructure. That is, instead of processing block among nodes in a serialized manner (i.e., process a block and then forward to the next peer), we design a mechanism that
supports transferring blocks of transaction data in a parallel manner among the nodes. Therefore, in BAASH, an individual node does not need to wait for the peers to complete a block mining process. To be more specific, each node performs the first step of the protocol (i.e., custom PoW) in parallel while gathering responses from fellow nodes about the second step of consensus (i.e., custom PBFT).
The transferring of any data block is handled by multiple processes, each of which handles the transfer to a particular node. For facilitating parallel blocks mining and consensus reaching process by having a handshake among the nodes, we implement an MPI wrapper for BAASH to be seamlessly integrated into HPC systems.

\subsubsection{Resilient distributed ledger}

In a large-scale HPC cluster, node failure is a norm rather than an exception.
This is especially true when MPI is used,
where a single faulty process can break down the entire job.
Therefore, to persist an updated ledger, 
we can leverage remote shared storage usually accessible through a parallel file system such as GPFS~\cite{gpfs_2002}.
The remote storage serves as a persistent medium in case of any node failure and can validate any block with PoW protocol with high difficulty (i.e., large nonce number). The reason for doing so is two-fold. \textit{First,} although remote storage is managed with an independent file system, 
we cannot trust the storage without the required verification. \textit{Second,} the block validation process will take place in the storage node only when the compute nodes fail to achieve consensus, which is rare and should not incur much performance overhead. 
In an MPI-powered blockchain system,
each node is managed by an individual process, also known as a rank in literature.
MPI is not fault-tolerant \textit{per se}:
the entire job execution can crash due to a single rank failure. 
Therefore, BAASH employs a fault-tolerant mechanism comprised of three components:
(i) a monitor that tracks each rank during the message exchanging, 
(ii) the \textit{atexit} routine, and (iii) exception handling. 
Besides, BAASH follows two initiatives in case of any kind of failure. First, the system tries to complete the block validation process with the help of remote storage. Second, if the remote storage fails, the job is automatically restarted so that the blockchain network can continue to make progress. 

This paper will present the new consensus protocol and analyze its complexity, describe the design and implementation of the system prototype using Python-MPI, and experimentally verify the system's effectiveness with more than two million transactions derived from micro-benchmarks on up to 500 nodes.
Experimental results show that our system prototype outperforms other blockchain systems:
$6\times$, $12\times$, and $75\times$ higher throughput than Hyperldeger, Ethereum, and Parity, respectively.
Besides, BAASH incurs orders of magnitude smaller latency and is almost $100\times$ faster than the conventional blockchain systems. 
In terms of resilience, our system prototype maintains high throughput under various degrees of faulty ranks.
To the best of our knowledge, BAASH is the first blockchain system working on and optimized for HPC.

To summarize, this paper makes the following scientific contributions: 
\begin{itemize}
    \item We design a set of consensus protocols to provide a double-layer of block validations with a parallel mechanism to make the consensus protocol less compute-intensive and less communication-intensive through MPI such that the highest possible throughput can be obtained without compromising the security in the proposed blockchain framework BAASH;
    
    \item We develop a parallel approach for processing the data blocks through MPI such that the system prototype becomes fully compatible with the mainstream shared-storage HPC infrastructure;
    
    \item We develop a resilience mechanism to keep the updated ledger in remote shared storage managed through a distributed file system that serves as a persistent medium in case of node failures;
    
    \item We implement a system prototype of the proposed consensus protocols and parallelization approaches with OpenMP and MPI; and
    
    \item We carry out an extensive evaluation of the system prototype with millions of transactions on an HPC platform and experimentally demonstrate the effectiveness of the proposed work by comparing it against the state-of-the-art blockchain systems.
\end{itemize}

The remainder of this paper is organized as follows.
In Section~\ref{sec:bg}, we discuss the background and some of the most recent related works. Section~\ref{sec:design} describes BAASH's design and its consensus protocols. 
Section~\ref{sec:implementation} presents the implementation details of BAASH and discusses some design trade-off. 
We report the experimental results and the lessons we learned during our development in Section~\ref{sec:eval}. 
Finally, we conclude this paper in Section~\ref{sec:conclusion}.

\section{Background}
\label{sec:bg}

\subsection{Preliminaries} 

A blockchain system is a distributed ledger that consists of multiple nodes, which are either fully or partially trustworthy. Some nodes can be compromised or intentionally injected in the network by the attackers that exhibit Byzantine behavior, even though the majority is honest. All the nodes maintain a shared replica of a blockchain or ledger (i.e., a set of shared blocks of transactions) and global states. Nodes are also responsible for generating, validating, and appending blocks with transactions in their local ledger as well as propagate them to the fellow nodes across the network to achieve global agreement about the validity of the block. The blockchain system has a special
data structure similar to a linked list that consists of a chain of blocks, where each block maintains the states and the history of transactions. It should be noted that a transaction in a block is the same as in the traditional database.

In order to make all the nodes fully trustworthy, several consensus protocols have been followed in the conventional blockchain systems such as proof-of-work (PoW), proof-of-stake (PoS), and practical byzantine fault tolerance (PBFT). In proof-of-work, each node in the network needs to solve a complex puzzle that needs a significant amount of computation, and the node that solves the puzzle is incentivized. In proof-of-stake, the creator of a new block is chosen in a deterministic way based on the wealth (i.e., stake) of the node, and there is no reward for mining a block. In PBFT, all of the nodes (partially trusted) within the system communicate with each other in order to reach an agreement about the state of the system through a majority. Therefore, nodes need to communicate with each other heavily, and not only have to prove that messages came from a specific peer node, but also need to verify that the message was not modified during transmission. 

There are two types of blockchain systems so far in the distributed ledger community, namely public and private blockchain systems. The private blockchain network is a permissioned network, where a node requires a special invitation. The network initiator then either verifies the node or prepares a set of rules to verify the node. This mechanism puts a restriction on who is allowed to participate in the network, and only in certain transactions. Once a node joins the network, it starts to play a role in maintaining the blockchain in a decentralized manner.

In a public blockchain network, anyone can join and participate in the network. The network follows an incentivizing mechanism to encourage more participants to join the network. One of the drawbacks of a public blockchain is, to achieve consensus, each node in a network must solve a complex, compute-intensive puzzle problem called proof-of-work (PoW) to ensure all data is synchronized and secured. This puzzle difficulty is needed to be increased along with the scaling of nodes to establish a trust that needs a substantial amount of computational power. Besides, the openness of public blockchain causes little privacy for transactions in the ledger and only supports a weak notion of security.

In conventional blockchains deployed to a cluster, each node holds a distinct copy of their ledger at the local disk.
Regardless of private (e.g., Hyperledger~\cite{hyperledger}) or public (e.g., Ethereum~\cite{ethereum}),
all existing blockchain systems assume that the underlying computer infrastructure is shared-nothing:
the memory subsystems and I/O subsystems are all independent on the participant nodes who are often connected through commodity networking such as Ethernet.

\subsection{Related Work}

At present, blockchain research focuses on various system perspectives. 
Algorand~\cite{crypto-algorand-Gilad:2017} proposes a technique to avoid the Sybil
attack and targeted attack. Bitcoin-NG~\cite{crypto-bitcoin-ng-Eyal:2016} increases throughput through a leader selection from each epoch who posts multiple blocks. Monoxide~\cite{Wang_nsdi19} distributes the workload for computation, storage, and memory for state representation among multiple zones to minimize the responsibility of full nodes. Sharding protocols~\cite{sharding-omniledger-kokoris2018, sharding-rapidchain-zamani2018} have been proposed to scale-out distributed ledger.
Hawk~\cite{kosba2016} is proposed to achieve transactional privacy from public blockchains.
 A blockchain protocol based on proof-of-stake called Ouroboros~\cite{Kiayias_2017}is proposed to provide stern security guarantees in blockchains and better efficiency than those blockchains based on proof-of-work (i.e., PoW). 
 Some recent works~\cite{sbft-gueta2018,honey-miller2016,bizcoin-kogias2016enhancing} propose techniques to optimize  Byzantine Fault Tolerance (BFT).
 Most recently, more consensus protocols, such as proof-of-reproducibility (PoR)~\cite{aalmamun_bigdata18} and proof-of-vote (PoV)~\cite{hpc-li2017proof} were designed either for in-memory architecture or to establish different security identity for network participants.

Being inspired by Hyperledger ~\cite{hyperledger}, Inkchain~\cite{inkchain} is designed as another permissioned blockchain solution that enables customization and enhancement in arbitrary scenarios. 
To improve reliability and achieve better fault tolerance, 
BigchainDB~\cite{bigchaindb} is built upon the Practical Byzantine Fault Tolerant (PBFT) and comes with blockchain properties (e.g., decentralization, immutability, owner-controlled assets) as well as database properties (e.g., high transaction rate, low latency, indexing and querying of structured data).
A 2-layer blockchain architecture (2LBC) is proposed in~\cite{aniello2017} to achieve stronger data integrity in distributed database systems based on a leader-rotation approach and proof-of-work.
Unfortunately, none of the aforementioned work addresses the underlying platform architecture other than shared-nothing clusters assumed by existing blockchain systems that could help us to bridge the gap between the HPC and blockchain technology.
The proposed blockchain framework presented by this paper, therefore,
for the first time,
showcases a practical parallel blockchain-like framework developed with MPI that allows us to leverage the decentralized mechanism in HPC systems.

In addition to the in-memory blockchain system~\cite{aalmamun_bigdata18}, we are well aware of recent advances in blockchain systems from the MPI and HPC communities. Various approaches~\cite{mpi-Chunduri-2018,mpi-Guo-2015,mpi-li2016, mpi-datampi-lu2014,mpi-consensus-buntinas2012,mpi-hive-chao2015, mpi-ppopp-Saillard2015} were proposed to improve or characterize the MPI features in order to design various solutions. However, all of these works are orthogonal to our work, and none of these aim to develop a new blockchain framework with MPI in order to facilitate the parallel processing in managing distributed ledgers. Therefore, the above-mentioned works can be merged into our work for further improvement in MPI specific packages.

\section{System Design}
\label{sec:design}

\subsection{Architecture Overview}

\begin{figure}[!t]
    \centering
    \includegraphics[width=75mm]{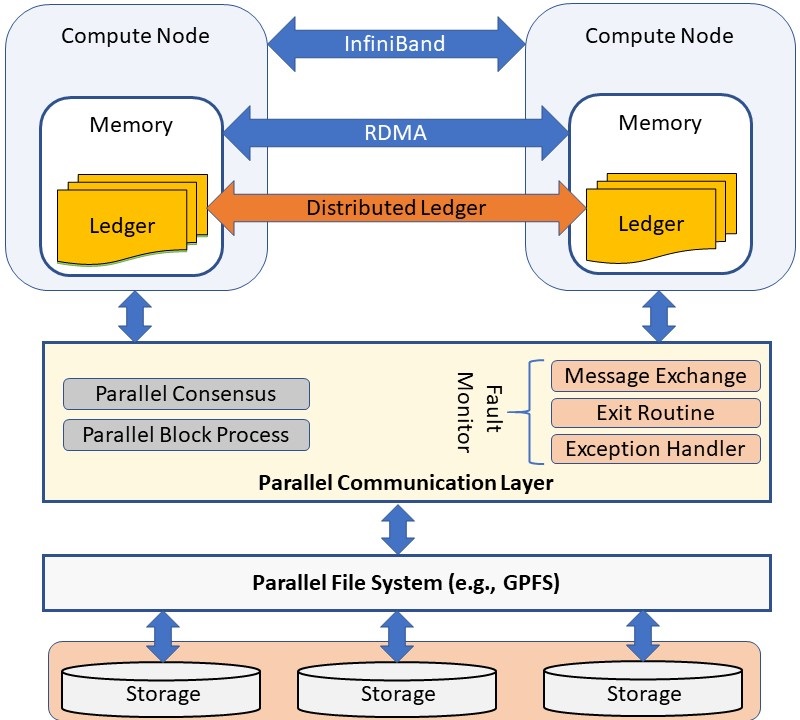}
    \caption{Overall architecture of BAASH on HPC systems. 
    }
    \label{fig:hpchain-ledger}
        \figspace
\end{figure}

Figure~\ref{fig:hpchain-ledger} illustrates the high-level overview of our envisioned distributed BAASH system, which is deployed to a typical HPC system. Note that some customized HPC systems may have node-local disks, although this paper assumes that the compute nodes are diskless.
For instance, a top-10 supercomputer called Cori~\cite{cori} located at Lawrence Berkeley National Laboratory does have local SSD storage, 
namely, the burst buffer.
However, the burst buffer is not a good candidate for ledgers because it is not designed for long-term data archival;
its only purpose is to alleviate the I/O pressure for those I/O-intensive applications by
caching the intermediate data locally.
Besides, because scientific applications do not time-share the resources,
the ledgers stored on the local storage (e.g., burst buffer) are usually purged after the job execution (for both performance and security reasons).
In other words, even if a ledger can be technically persisted to local disk in some scenarios,
that persistence is not permanent,
which motivates us to come up with a secondary ledger and validator on the remote storage.
Specifically, four key modules of our envisioned system are highlighted in Figure~\ref{fig:hpchain-ledger}:
a distributed ledger, parallel communication layer, a fault monitor, and a remote persistent ledger.
We will discuss each of them in more detail in the following.

\subsubsection{Distributed Ledger}

The first module is a distributed ledger implementation optimized for high-performance interconnect (e.g., InfiniBand) and protocols (RDMA) across compute nodes.
In particular, the high-performance hardware should be able to significantly speed up communication-intensive consensus algorithms such as practical byzantine fault tolerance (PBFT)~\cite{jsousa_dsn18} widely used in permissioned blockchains (e.g., Hyperledger~\cite{hyperledger}). 
We are specifically interested in permissioned blockchains because the HPC environment for scientific applications allows only authenticated participants 
(as opposed to permissionless blockchain systems, e.g., Ethereum~\cite{ethereum}, which are open to the general public).

\subsubsection{Parallel Communication Layer}

The second module is the communication layer between the diskless compute nodes and the remote persistent storage.
This layer has three purposes:
(i) It injects the parallel communication mechanism for transferring blocks with transactions,
(ii) It facilitates parallel block processing and achieving consensus by having handshake among the nodes and, (iii) Compute nodes can persist the ledgers in the remote storage in a parallel manner. 
The key question is how to persist the in-memory ledger into the remote parallel file system.
The persistence cannot occur for every transaction as the enormous I/Os would become the performance bottleneck of the application. Therefore, we leverage a hash table to solve this problem, where we store the hash ids of the blocks persisted in the remote storage. We only enable a compute node to persist a block in the remote storage, when it does not find the hash of a newly validated block in the hash table. 
That is, the compute node who will first be able to validate the block will only get access to the remote storage to persist it.

\subsubsection{Fault Monitor}

The third module is a fault tolerance mechanism implemented with three inevitable checkpoints: 
\begin{enumerate}
    \item Message exchange among peers: Check the communication status between the sender and receiver during each message exchange. 
    \item Exit routine of each process: An exit routine that checks the status of each node when the node finishes a validation process.  
    \item Exception handler: Set checkpoints around the node validation process to catch any unexpected exceptions.
\end{enumerate}

A monitoring process is designed to intercept the three aforementioned checkpoints. If any of the checkpoints raises an alert during the block validation process, the remote storage comes forward first to complete the block validation process. That is, the block is then continued to be validated against the ledger replicated in the remote storage. In the case of remote storage failure, the whole validation process is restarted automatically, which is the worst case and very rare to happen. 

\subsubsection{Remote Storage}

The fourth module focuses on persisting in distributed remote storage accessed through the parallel file system (i.e., GPFS).
Because all compute nodes fetch and update the ledger only in volatile memory (or, in persistent local storage but with a short lifetime---purged after the job is complete),
there has to be at least one permanent, persistent storage (which cannot be compromised) to store back the ledger replicas in case of catastrophes (e.g., more than 50\% of compute nodes crash and lose their ledgers).
Note that the memory-only compute nodes are not necessarily less reliable than those with persistent storage;
yet, the data stored on memory would get lost if the process which initiates the memory is killed.

\subsection{High-Performance Blockchain Enhanced (HyBE) Protocol}

In order to overcome the limitations of the conventional protocols (i.e., PoW and PBFT), as a co-design of the proposed BAASH blockchain framework for scientific applications, we propose a set of consensus protocols that enable the compute nodes to process the blocks in a parallel manner and consequently, make the proposed framework fully compatible with the HPC infrastructure. Besides, the proposed mechanisms of the validation process in the newly designed protocols do not degrade the performance along with the scaling of nodes, which in turn makes the BAASH highly scalable.  

The first protocol, \textit{Persist-in-Storage}, aims to persist blocks from memory of the compute nodes into remote storage through a parallel file system (i.e., GPFS) to back up a ledger replica only in the observation that all ledgers on compute nodes are essentially volatile.
The second protocol, \textit{HyBE Consensus}, further extends the first protocol by taking the ingredients of both the PoW and PBFT protocols to achieve consensus when a new block is submitted.

\begin{algorithm}
\floatname{algorithm}{Protocol}
    \caption{Persist blocks to remote storage}
    \label{alg:persistent}
    \begin{algorithmic}[1]
        \Require 
        Compute nodes $C$ where the $i$-th node is $C^i$; 
        $C^i_B$ the local blockchain on $C^i$;
        a newly mined block $b$; 
        remote storage $R$;
        $H$ the hash table that contains hashes of blocks stored in remote storage;
        $R_B$ the blockchain copy on the storage;
        \Ensure Validate $b$ and persist it to $R$
        \Function{Persist-in-Storage}{$b$, $C$, $R$, $H$}
        \For {$C_i \in C$}
        \If {$b$ is valid with $C_B^i$} 
        \State $C^i_B \gets C^i_B \cup b$ \Comment{b is appended to the chain}
        \If {$b \not\in H$} \Comment{Only one look-up is needed}
        \State $R_B \gets R_B \cup b$ 
        \EndIf
        
        \EndIf
        \EndFor
        \EndFunction
    \end{algorithmic}
\end{algorithm}

\begin{algorithm}
\floatname{algorithm}{Protocol}
    \caption{All compute nodes reach consensus}
    \label{alg:consensus}
    \begin{algorithmic}[1]
        \Require 
        Compute nodes $C$ where the $i$-th node is $C^i$; 
        $C^i_B$ the local blockchain on $C^i$;
        a new block $b$; 
        remote storage $R$;
        $R_B$ the blockchain copy on the storage;
        \Ensure At least 50\% compute node list $agreedNodes$ who validate $b$ both with local blockchain and with remote persistent ledger $S_B$.
        
        \Function{HyBE-Consensus}{$b$, $C$, $S$}
               \For {$C_i \in C$} \Comment{in parallel}
              \If {$b$ finds valid hash with $C_B^i$} 
                \For {$C_j \in C$} \Comment{in parallel}
                  \If {$C_i$ agrees with $C_j$}   
                  \State $agreedNodes  \gets agreedNodes \cup C_i$ 
                  \EndIf
                 \EndFor          
              \EndIf
            \EndFor
         
        \If{$| agreedNodes | <= \frac{N}{2}$} 
          \If {$b$ finds valid hash with $R_B$}
          \State 
          $agreedNodes  \gets agreedNodes \cup R$ 
          \EndIf
         \Else  
                 \For {$C_i \in C$}  \Comment{Protocol 1}
                    \State $C_B^i \gets C^i_B \cup b$ \Comment{Store on compute node}
                \EndFor
            \State $R_B \gets R_B \cup b$ \Comment{Store on storage node}
        \EndIf
        \EndFunction
    \end{algorithmic}
\end{algorithm}

\subsubsection{Persist-in-Storage}

The protocol is described in Protocol~\ref{alg:persistent}.
In essence, a block is appended in parallel to the blockchain in local memory and persisted to a parallel file system (e.g., GPFS),
as shown at line 4 and line 6.
Doing so adds one more layer of reliability to the data on volatile memory. 
However, while attaining high reliability, 
the system should not exhibit significant overhead. 
This is addressed in line 5, where we only need one look-up to check in the hash table that holds the hashes of all blocks stored in the storage node. This prevents touching the shared storage more than once when other nodes try to store the same block again.
Although the theoretical time complexity of this protocol is $O(|C|)$ and $|C|$ could be a fairly large number (e.g., tens of thousands of cores in leading-class supercomputers~\cite{cori}).
We will demonstrate the effectiveness of the protocol in the evaluation section.
The correctness of Protocol~\ref{alg:persistent} is obvious because the only change to the original PoW consensus is the data persistence,
which has nothing to do with the agreement between the compute nodes. The main goal of this first protocol is to help the HyBE protocol (i.e., Protocol~\ref{alg:consensus}) in continuing the block validation process. 

\subsubsection{HyBE Consensus}

The second part of the proposed consensus protocol called High-performance Blockchain Enhanced Protocol (HyBE) consists of two phases, as shown on Protocol~\ref{alg:consensus}. The first phase is conducted in parallel in all compute nodes, and it utilizes the concept of PoW but with reduced difficulty (i.e., low nonce), as shown in Line 3. The second phase uses the modified version of PBFT (i.e., parallel message passing) to handshake with fellow nodes in a parallel manner to reach an agreement about the validation of a block, as shown in Line 5. In this modified version of PBFT, we replace the concept of serialized message passing technique with our proposed parallel approach,
which makes the block validation process significantly faster. Together these two phases make all the participants in the network fully trustworthy. However, to inject resiliency in the proposed prototype, an extra layer of reliability has been added that leverages the remote shared storage as a backup validator in case of any kind of catastrophe (i.e., line 11). That is, if more than 50\% compute nodes fail to provide consensus, the remote storage then comes forward to validate the block, as shown in Line 12. The time complexity of this protocol is $O(|C|^2)$,
which is on par with the existing PBFT algorithm. However, in practice, the complexity of the protocol is reduced to $O(|C|)$, as our proposed mechanism through MPI helps us to avoid another $O(|C|)$ iterations to get consensus from the fellow nodes. Therefore, we could achieve the consensus from the fellow nodes with comparably fewer iterations than the conventional PBFT.

\section{Implementation}
\label{sec:implementation}
We have implemented a prototype system of the proposed blockchain architecture and consensus protocols with Python-MPI. MPI is a communication protocol for programming parallel computers. In MPI, both point-to-point (sends, receives) and collective (broadcasts) communication are supported. More specifically, MPI is a message-passing application programmer interface, together with protocol and semantic specifications for how its features must behave in any implementation~\cite{GROPP1996}. We use the mpi4py package~\cite{mpi4py} for leveraging MPI with Python. This MPI package for Python provides bindings of the MPI standard for the Python programming language, allowing any Python program to exploit multiple processors across machines. Typically, for maximum performance, each CPU (or core in a multi-core machine) will be assigned one single process or a distinct \textit{rank}.

\begin{figure}[!t]
    \centering
    \includegraphics[width=75mm]{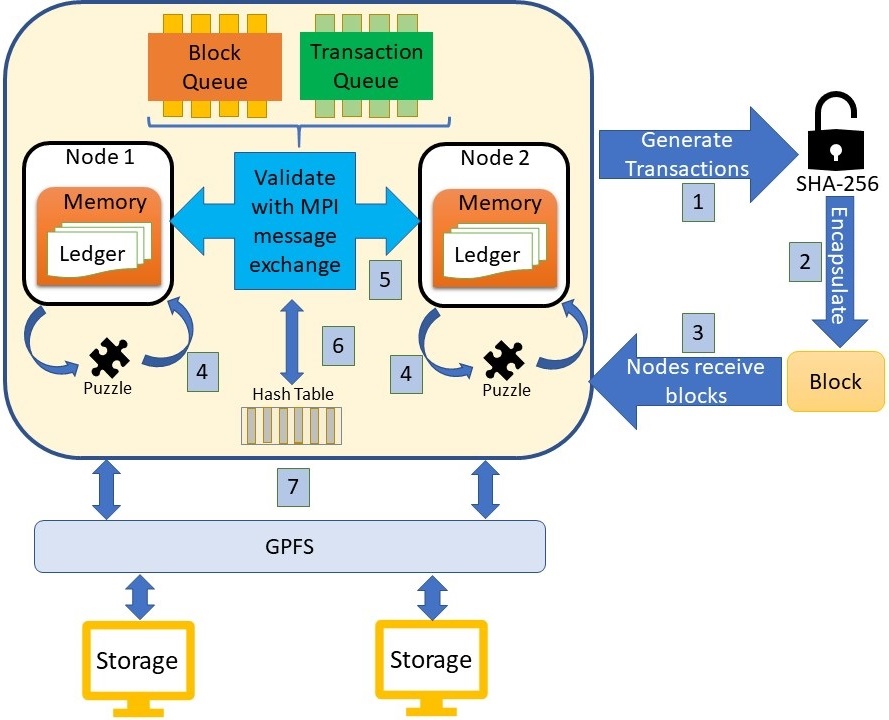}
    \caption{BAASH implementation with MPI and shared storage.
    }
    \label{fig:hpchain-impl}
        \figspace
\end{figure}

At this point, we only release the very core modules of the prototype.
Some complementary components and plug-ins are still being tested with more edge cases and will be released once they are stable enough.
Figure~\ref{fig:hpchain-impl} illustrates the overview of the implemented architecture of the proposed parallel blockchain on MPI. The prototype system has been deployed to 500 cores on an HPC cluster.

As shown in Figure~\ref{fig:hpchain-impl}, new transactions generated by the nodes are first hashed using SHA-256~\cite{sha256} (step 1). 
Then, the transactions are encapsulated in a block by the respective nodes (step 2) before broadcasted among the other nodes (step 3). Afterward, the blocks are validated first in compute nodes (step 4) using custom PoW. Then, based on custom PBFT, another step of the validation process is resumed to receive a vote for consensus from fellow compute nodes (step 5). However, if more than 50\% nodes fail to provide consensus, remote storage then participates in the block validation process (step 6). Finally, if the block is validated successfully, BAASH decides to append the block both in the in-memory ledger of the compute nodes and in the remote storage (step 7) through a parallel file system (i.e., GPFS) by checking the hash table first to ensure that the block is not already stored.

\subsection{Worker Nodes}

In BAASH, nodes are responsible for spawning transactions in random orders with a fixed time interval. The transactions are then appended to the transaction queue, which is discussed in detail in~\S\ref{sec:queue}. When the created transaction queue reaches a limit, the nodes encapsulate them (i.e., transactions) in a block. When a block is encapsulated with a number of transactions, it is then pushed into a queue (discussed in ~\S\ref{sec:queue}) before it is broadcasted in the network. 

The nodes are also responsible for validating the blocks and sending consensus to the fellow nodes, which is discussed in~\S\ref{sec:consensus}. A node will store a block both in its local in-memory as well as in the remote storage after validation. The format of storing data in a compute node and in remote storage is explained in ~\S\ref{sec:datamodel}. Each node communicates with each other through our parallel mechanism implemented with MPI discussed in~\S\ref{sec:mpi-interface}. The communication between compute nodes and the remote storage is managed through a parallel file system (e.g., GPFS).

\subsection{Parallel Processing and Broadcast Module}

We develop a module with a parallel mechanism that enables the BAASH to transfer and validate blocks and transactions among the fellow compute nodes in a parallel manner. Our module leverages point-to-point service to carry out communication between two fellow nodes, whereas the handshake part (i.e., receiving consensus from other nodes) is managed through collective communication service. As BAASH is built with the mixture of modified version of PoW (i.e., moderate nonce) and PBFT (i.e., less communication-intensive) (discussed in~\S\ref{sec:consensus}) to make the validation process less computation and communication-intensive, the proposed parallel communication mechanism provided by our designed module helps us to convert the conventional PBFT so that it requires significantly less communication among the peers.

A broadcast module is also developed with the MPI programming interface to manage the parallel propagation of the blocks and transactions among all the compute nodes. For facilitating the parallel transfer of the blocks and transactions, we exploit two dynamic hash tables for queuing blocks and transactions. The queues are arbitrarily accessible so that the blocks and transactions can be transferred in parallel to multiple nodes. It should be noted that, in the present blockchain systems, all the blocks and transactions are transferred among the compute nodes using the serialized queue. This serialization mechanism is one of the critical issues in performance for the HPC systems and causes significant communication delay and latency.

\label{sec:mpi-interface}

\subsection{Block and Transaction Queue}
\label{sec:queue}
All the newly created blocks by a node are pushed to an outgoing FIFO queue before they are broadcasted to the other nodes. At a fixed time interval, a block pops out from the queue periodically and is transferred to the connected nodes. To keep the incoming blocks in proper order, each node has an incoming FIFO queue, which stores the incoming blocks from the connected nodes. A block is popped out from the incoming FIFO queue as soon as the node completes the appending process of the previous block. It should be noted that when a node creates a new block, it also pushes the newly created blocks in its incoming queue so that the node can participate in the validation process for that block along with the other fellow compute nodes.  

Similarly, the newly created or received transactions are pushed into a FIFO queue to start the block encapsulating process. In BAASH, we can adjust the queue size according to our demand. For instance, for all the experiments presented in this paper, each block consists of 4 transactions on average. Therefore, the block encapsulation process is triggered as soon as the queue size reaches 4. 
This is because we want to keep our experimental results compared with the benchmarks~\cite{blockbench_sigmod17}.

It could be argued that a queue might not deliver data at a sufficient rate to feed the network because a queue is a linear data structure that can hardly be parallelized for the sake of scalability.
This is alleviated by the following two approaches in our implementation.
First, we adjust the time interval larger than the latency for the queue to pop an element. In doing so, the overhead from the queue itself is masked completely.
Second, we implement the queue using a loosely-coupled linked-lists such that the queue can be split and reconstructed arbitrarily.

\begin{figure}[!t]
    \centering
    \includegraphics[width=75mm]{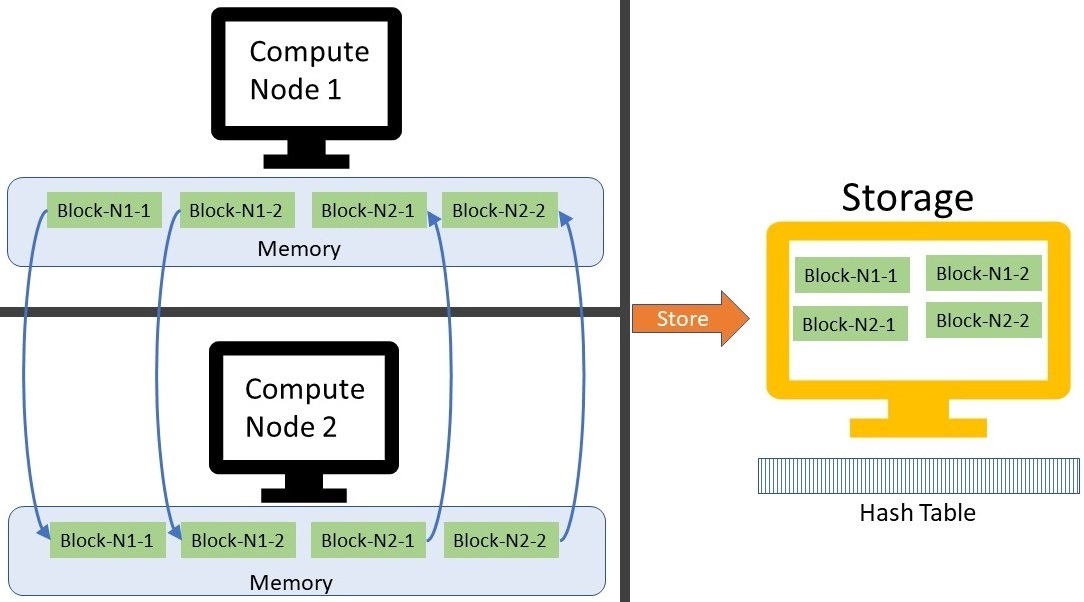}
    \caption{Data structure of BAASH's distributed ledger. The chain of blocks created and appended by various nodes are stored in the persistent storage.
    }
    \label{fig:hpchain-ds}
        \figspace
\end{figure}

\subsection{Data Models and Storage}
\label{sec:datamodel}
The data structure for the proposed BAASH ledger is a linked list and stored in a row-wise table where each tuple corresponds a block, which references a list transaction records stored in another table,
just like the conventional relational database tables. 
As shown in Table~\ref{tbl:block} and Table~\ref{tbl:transaction}, 
all properties of a block (e.g., block ID, parent block hash, transactions list, and time stamp) and all properties of a transaction (e.g., transaction ID, sender ID, receiver ID, amount, time stamp) are encapsulated respectively in a \texttt{Block} object and in a \texttt{Transaction} object at runtime.

Figure~\ref{fig:hpchain-ds} shows a concrete example of the structure to store the blockchain on two specific nodes (i.e., Node 1 and Node 2). For instance, under Node 1, there are four blocks, among which two of them are created by itself (i.e., Block-N1-1 and Block-N1-2), and the rest of the blocks are from Node 2 (i.e., Block-N2-1, Block-N2-2). Each node is also connected with remote storage through a parallel file system (i.e., GPFS). Each block is appended in the blockchain of the remote storage by a node after the validation process is completed. However, the appending process in remote storage for a particular block only happens once and only by the node who can solve the validation process of the block first. We use a hash table to keep track of the hashes of the blocks in the remote storage. Therefore, when a node attempts to store a block in the storage node, it first looks up quickly in the hash table to check whether the block is already stored in the storage. If the block hash is already in the hash table, the respective node does not attempt to access the remote storage further. It should be noted that each node in the network has a copy of the hash table, which is periodically updated with a fixed time interval.

\begin{table}[!t]
    \caption{Block Structure}
    \centering
    \begin{tabular}{ lll }
    \toprule
        \textbf{Field} & \textbf{Description}    \\
    \midrule
    blockID & Unique Identifier of the Block \\    \hline
    
    parentBlockHash &  Hash Value of Parent Block \\    \hline
   
    creationTime & Creation Time of Current Block \\    \hline
     receiveTime & Receiving Time of Current Block \\    \hline
     txnCounter & Number of Transactions in a Block \\    \hline
    txnList & The List of Transactions in a Block \\ \hline
    puzzleDifficulty & Number of nonce to set difficulty \\ 
    \bottomrule
    \end{tabular}
    \figspace
    \label{tbl:block}
\end{table}

\begin{table}[!t]
    \caption{Transaction Structure}
    \centering
    \begin{tabular}{ lll }
    \toprule
        \textbf{Field} & \textbf{Description}    \\
    \midrule
    txnID & Unique Identifier of the Transaction \\    \hline
    
    senderID & Sender ID (Node ID) of Transaction \\    \hline
    receiverID & Receiver ID (Node ID) of Transaction \\    \hline
    creationTime & Creation Time of Transaction \\    \hline
     receiveTime & Receiving Time of Transaction \\    \hline
     txnAmount & Amount of Transaction\\    
    
    \bottomrule
    \end{tabular}
    \figspace
    \label{tbl:transaction}
\end{table}

\subsection{Consensus Protocol}

The consensus protocol between compute nodes, as a building block of the proposed HyBE consensus, shares the same essence of the conventional PoW but simplifies the compute-intensive puzzle taken by Bitcoin~\cite{bitcoin}. Besides, it injects another layer of reliability using our parallel message passing mechanism through MPI to get the agreement from fellow nodes to reach consensus (i.e., modified PBFT). 

\begin{figure}[!t]
    \centering
    \includegraphics[width=75mm]{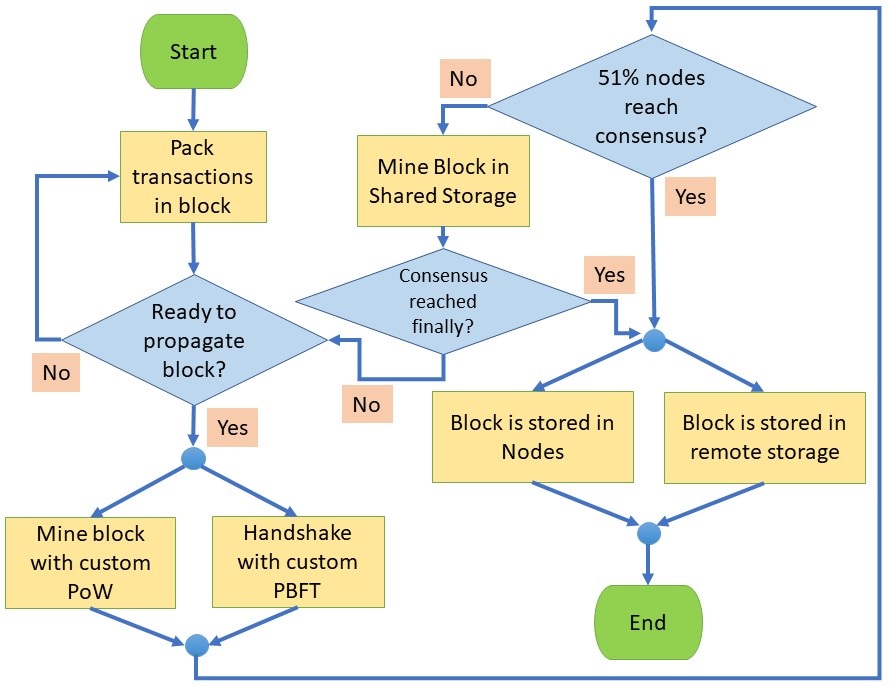}
    \caption{Workflow of HyBE protocol.
    }
    \label{fig:consensus-workflow}
        \figspace
\end{figure}

The workflow of the proposed HyBE protocol in the context of BAASH blockchain is illustrated in Figure \ref{fig:consensus-workflow}.
Our current consensus protocol implementation is composed of four steps.
First, all the newly created transactions generated in a node are added to that node as well as propagated to other peers and eventually packed into a block by all the peers in the network for the preparation of starting the mining process.
Second, all the compute nodes will first validate the block using customized PoW (i.e., with moderate nonce).
Third, all compute nodes will handshake with each other using the customized version of PBFT, which is specially crafted to handle parallel communication among large scale of nodes. If more than 50\% nodes agree about the validity of the block, it is then appended both in the memory ledger of compute nodes as well as in the remote storage. Finally, if at least 51\% nodes fail to provide consensus, the remote storage then comes forward to validate the block, and if it (i.e., block) does not get validated in this round, it will be pushed to a waiting queue for restarting the validation process again.

\subsection{Fault Tolerance Monitor}
We develop a fault tolerance module to manage the node failure with three checkpoints. In the first checkpoint, each peer is monitored during the phase of the message exchange to ensure successful communication. An exit routine is registered that checks the status of each node when it (i.e., node) finishes the block validation process. The second checkpoint keeps monitoring this exit routine for graceful termination. The final checkpoint
watches over all kinds of exceptions thrown by the nodes during the entire execution. If any of the checkpoints raise any alert, the block validation process continues with the help of the remote storage that holds the most accurate ledger. 

\label{sec:consensus}

\section{Evaluation}
\label{sec:eval}

\subsection{Experimental Setup}

\textbf{Testbed.}
All experiments are carried out on a high-performance computing cluster comprised of 58 nodes interconnected with FDR InfiniBand. 
Each node is equipped with an Intel Core-i7 2.6 GHz 32-core CPU along with 296 GB 2400 MHz DDR4 memory.
There is no local disk on compute nodes, which is a typical configuration on HPC systems;
yet, a remote 2.1 PB storage system is available managed by GPFS~\cite{gpfs_2002}.
Each node is installed with Ubuntu 16.04, Python 3.7.0, NumPy 1.15.4, mpi4py v2.0.0, and mpich2 v1.4.1. 
We deploy our system prototype on up to 500 cores and report the average results unless otherwise noted.

\textbf{Workload.}
For micro-benchmarks, we use YCSB workload~\cite{ycsb} commonly used for transaction evaluation. 
The reason why we did not choose popular cryptocurrency data (e.g., Ethereum ERC20 used in~\cite{Wang_nsdi19}) was that the targeting data of HPC blockchains are not necessarily in financial domains.
In our system prototype, each block contains four transactions, and we deploy more than two million transactions (2,013,590) in all of the experiments. 

\textbf{Evaluation Metrics.}
We evaluate our system prototype in the following four metrics:
\begin{enumerate}

\item Throughput: measured as the number of successful
transactions per second. 

\item Latency: measured as the response time per transaction. 

\item Scalability: measured as the changes in throughput
and latency when increasing the number of compute nodes.

\item Fault tolerance: measured as to how the ledger's validity and throughput vary when some nodes fail.
\end{enumerate}

\textbf{Systems for Comparison.}
We compare Ethereum~\cite{ethereum}, Parity~\cite{parity}, and Hyperledger Fabric~\cite{hyperledger} with the proposed BAASH framework, whose original codename is HPC-blockchain, 
or more commonly known as HPChain. 

\subsection{Throughput and Latency}

\begin{figure}[!t]
    \centering
    \includegraphics[width=75mm]{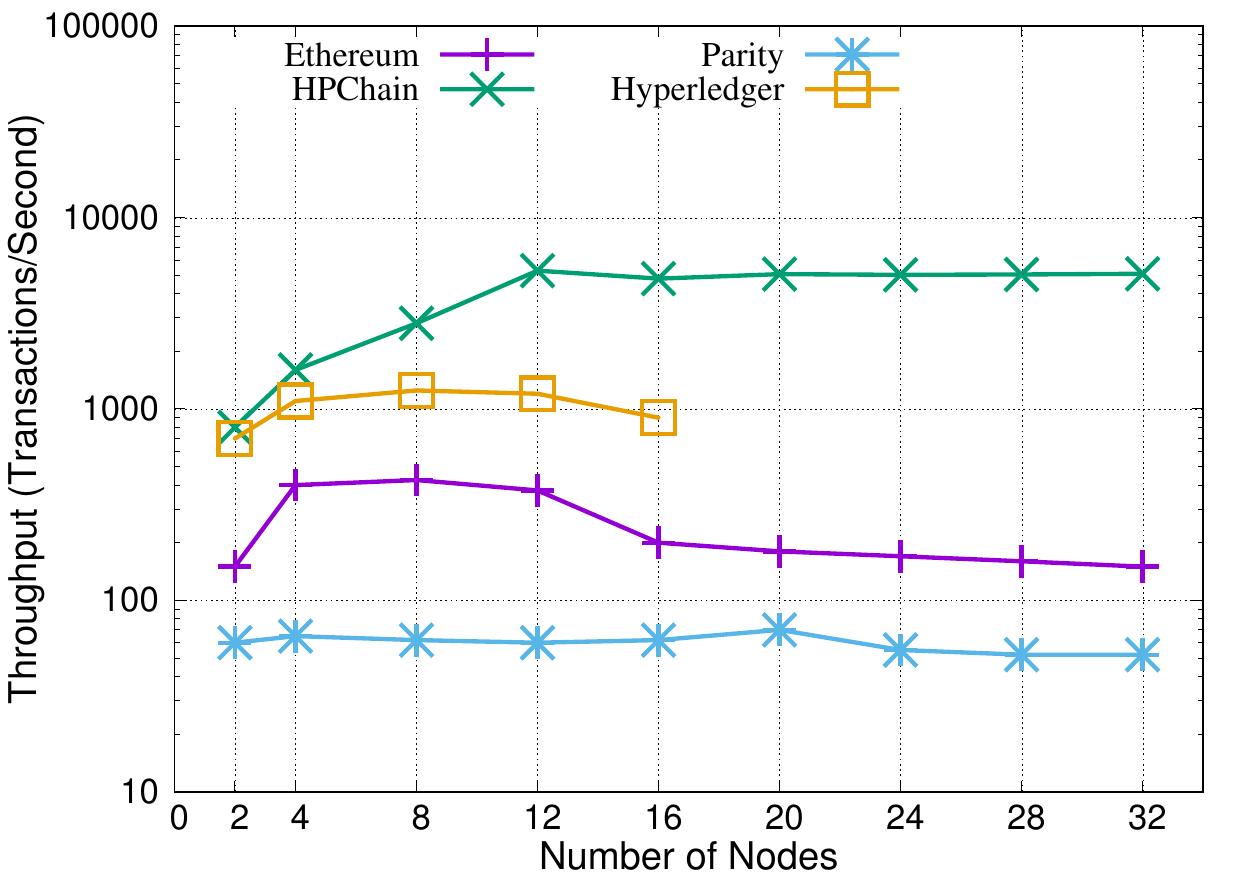}
    \caption{Transactions generated per second. Each node processes the same number of blocks and transactions.
    }
    \label{fig:txn-throughput-32-node}
        \figspace
\end{figure}

\begin{figure}[!t]
    \centering
    \includegraphics[width=75mm]{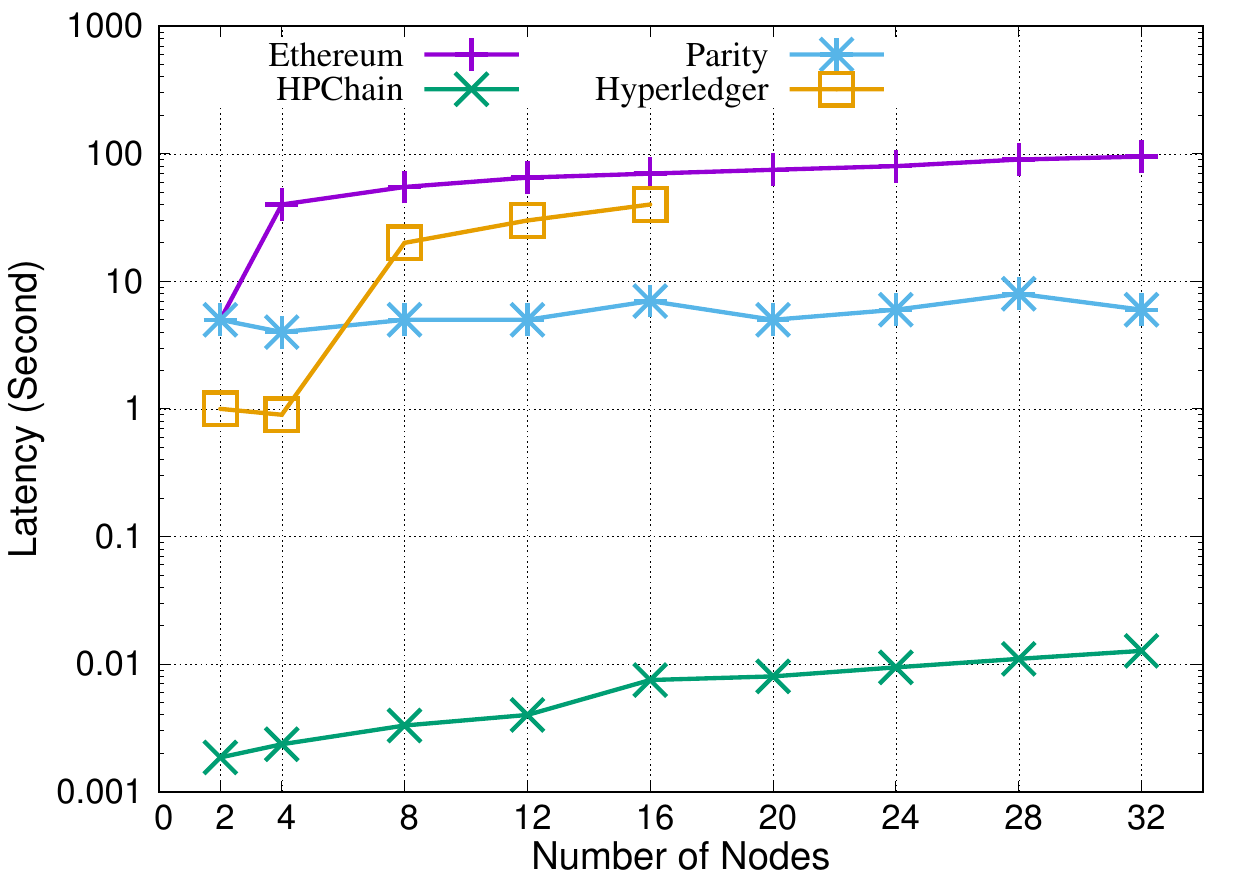}
    \caption{Latency of transaction processing
    }
    \label{fig:txn-latency-32-node}
        \figspace
\end{figure}

Each node sends 900 transactions per second. Figure~\ref{fig:txn-throughput-32-node} and Figure~\ref{fig:txn-latency-32-node} show the throughput and latency with varying node scales.
In terms of throughput,
HPChain outperforms other systems at all scales. 
Specifically, it has up to $6\times$, $12\times$, and $75\times$ higher throughput than Hyperldeger, Ethereum, and Parity, respectively, on 16 nodes. 
Thanks to the proposed consensus protocols that are optimized for HPC systems,
HPChain's throughput is not degraded at larger scales beyond 16 nodes, while both Ehtereum and Parity show some degradation.
Worse yet, Hyperledger cannot scale beyond 16 nodes at all.

HPChain also incurs significantly lower latency than other major blockchain systems:
$1000\times$, $400\times$, and $5000\times$ less than Hyperledger, Parity, and Ethereum, respectively. 
As shown in Figure~\ref{fig:txn-latency-32-node}, HPChain incurs only less than 20 milliseconds at 32 nodes because of the parallel communication technique along with the tricky mixture of protocols (i.e., modified PBFT with low difficulty in PoW) that helps in achieving consensus faster while keeping the message channel free among the nodes.

\begin{figure}[!t]
    \centering
    \includegraphics[width=75mm]{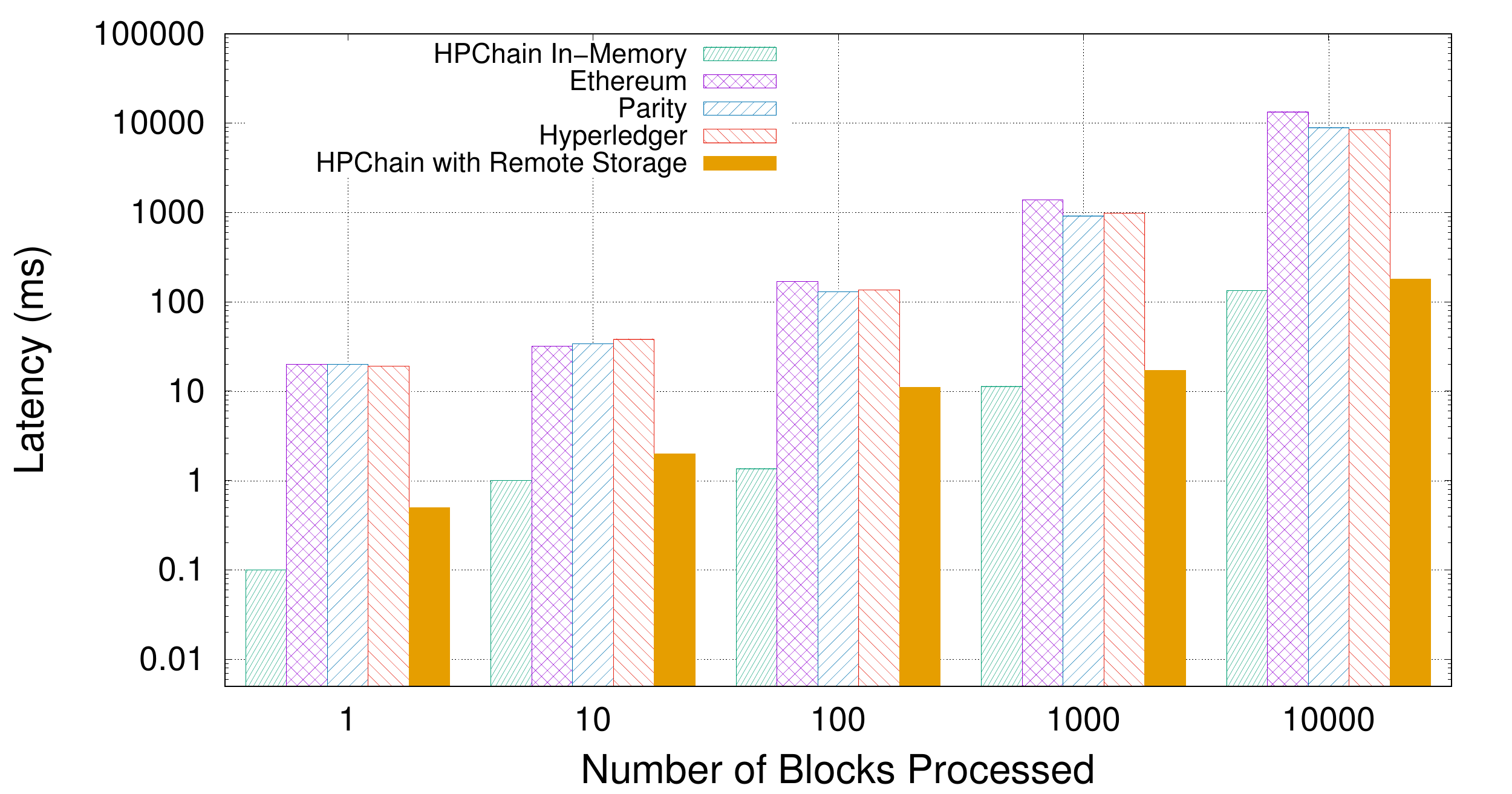}
    \caption{Time for processing different numbers of blocks.
    }
    \label{fig:latency-block-process}
        \figspace
\end{figure}

To further study the latency incurred by all the systems, another experiment is carried out with a micro-benchmark workload on two nodes where each block contains three transactions on average. The transaction transfers a value from one random account
to another random account. Figure~\ref{fig:latency-block-process} reports HPChain incurs the lowest latency (i.e., almost only 100 ms), which is significantly lower (i.e., almost $100\times$) compared to others. This further proves the strength of the proposed mechanism in HPChain. We also analyze HPChain's performance with remote storage validation. We intentionally enable HPChain to access the remote ledger even when the blocks were processed successfully with the compute nodes. We observe that HPChain, with both in-memory and remote storage validation, incurs negligible overhead compared to the HPChain with in-memory support only. This is because the remote ledger is designed to be accessed only once and only when 51\% compute nodes fail to provide consensus. 

\subsection{Scalability}

In this section, we report HPChain's PoW performance at
various scales up to 500 nodes. 
 By scalability, we want to measure the throughput (i.e., number of transactions or number of processed blocks) per second. Bitcoin block can be mined within every ten minutes ~\cite{bitcointy}, and this is because of the extensive increment in puzzle difficulty along with the scalability of nodes, which makes the Bitcoin unscalable~\cite{kzhang_icdcs18}.
In fact, it is well-accepted that existing PoW, PoS, and PBFT blockchain protocols are not scalable yet~\cite{kzhang_icdcs18}. 

HPChain does not use high puzzle difficulty on the scaling of nodes to increase reliability. Besides, we argue that HPChain is the first blockchain framework with a parallel mechanism (as per our knowledge) specially crafted for HPC systems that supports parallel processing. Therefore, at this stage, the scalability of HPChain is not compared with other conventional blockchain systems. 

\begin{figure}[!t]
    \centering
    \includegraphics[width=75mm]{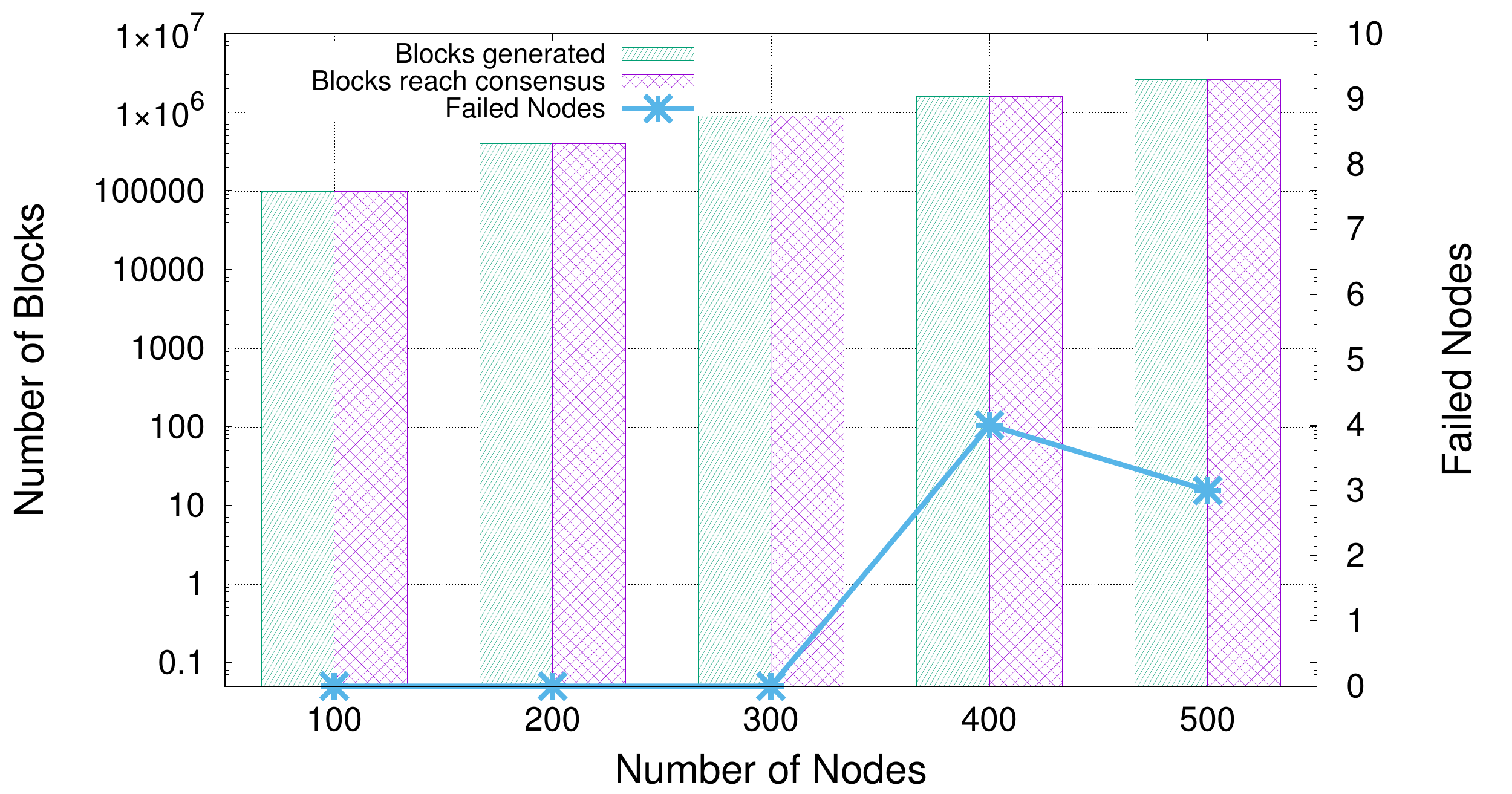}
    \caption{Blocks reaching consensus.
    99\%+ blocks reach consensus in HPChain without compromising the performance in throughput at all scales.
    }
    \label{fig:compare-block-consensus-scalability}
        \figspace
\end{figure}

Figure~\ref{fig:compare-block-consensus-scalability} reports the number of blocks that reach consensus in HPChain at different scales ranging from 100 to 500 nodes. 
We notice that in all scales, all the generated blocks can reach consensus even with node failures. However, the possibility of node failure is negligible. That is, only four nodes on 400 scale (i.e., 1\%) and three nodes on 500 scale (i.e., 0.6\%) tend to fail during the experiment. 
A blockchain can work correctly as long as at least 51\% of the nodes are not compromised; Therefore, the handful of failed nodes is negligible.

\begin{figure}[!t]
    \centering
    \includegraphics[width=75mm]{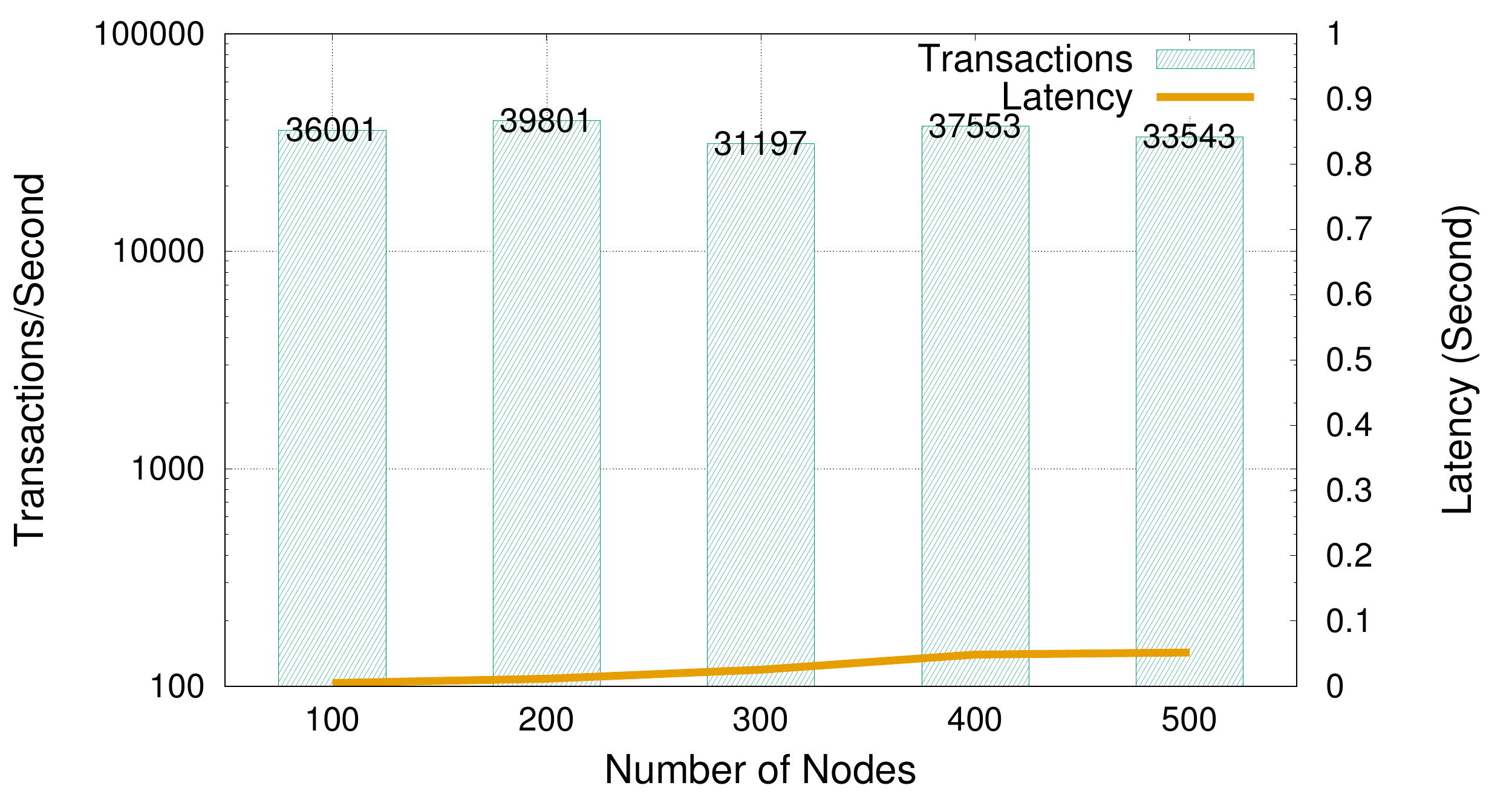}
    \caption{Transaction throughput is not significantly influenced by the scaling of nodes.
    }
    \label{fig:compare-txn-scalability}
        \figspace
\end{figure}

We then further investigate the scalability of HPChain in terms of transaction throughput. Figure~\ref{fig:compare-txn-scalability} shows the result at different scales ranging from 100 to 500 nodes. HPChain shows an insignificant gap in throughput and incurs minor latency (i.e., below 0.1 seconds on up to 500 nodes). This means, in HPChain, performance does not get affected by the scaling of nodes. Keeping constant performance, along with scaling, is one of the major issues in blockchain systems, as the present systems are not scalable~\cite{kzhang_icdcs18}. As shown in Figure~\ref{fig:txn-throughput-32-node}, while in other systems throughput tends to fall just after 16 nodes, the HPChain produces significantly higher throughput while keeping it almost at a constant rate even at the large scale.

\subsection{Fault Tolerance and Reliability}

\begin{figure}[!t]
    \centering
    \includegraphics[width=75mm]{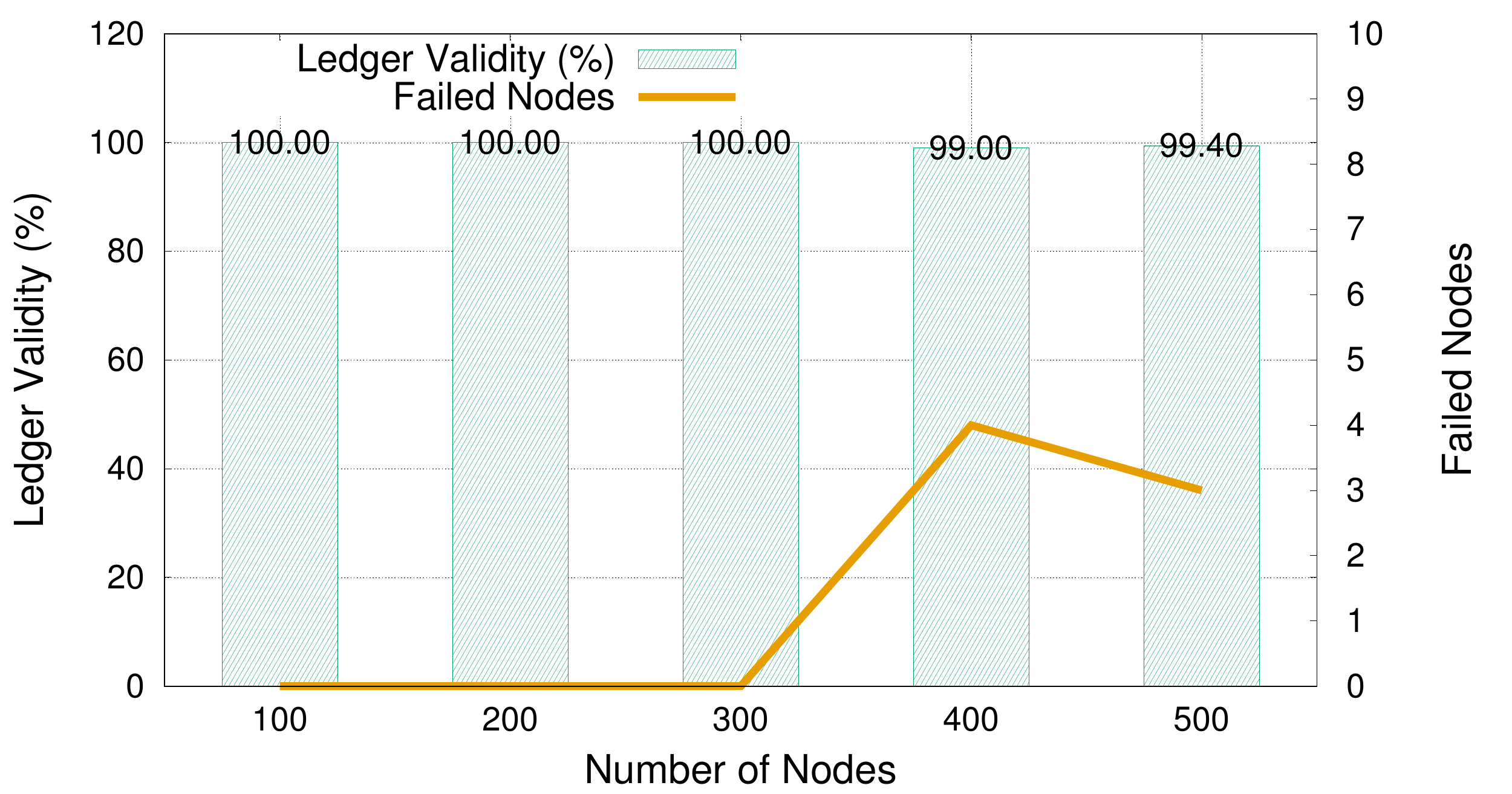}
    \caption{Ledger validity in HPChain:
    99\%+ nodes keep a valid ledger at all scales.
    }
    \label{fig:ledger-validity}
        \figspace
\end{figure}

We measure the reliability of HPChain on up to 500 nodes and check how many nodes can hold 100\% valid ledgers. We also keep track of how much node failure occurs during this experiment. Figure~\ref{fig:ledger-validity} shows that in all scales, at least 99\% nodes keep the valid ledger. We find 100\% nodes hold valid ledger on up to 300 nodes. However, HPChain tends to show negligible node failure (i.e., at most 1\% nodes) in larger scales (i.e., 400 and 500 nodes). It should be noted that a blockchain can work correctly as long as at least 51\% of the nodes are not compromised. By node failure, we mean some of the nodes remain offline due to the hardware insufficiency.

\begin{figure}[!t]
    \centering
    \includegraphics[width=75mm]{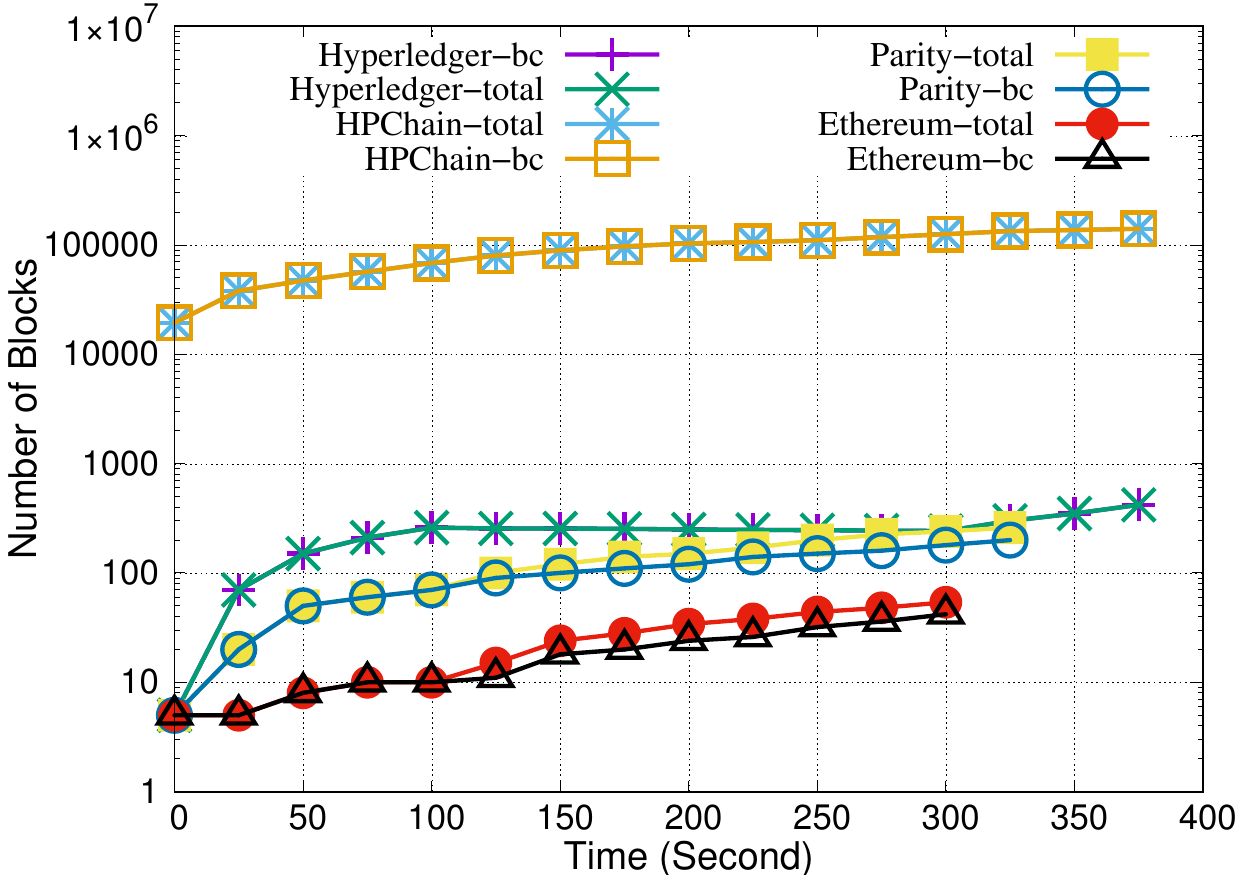}
    \caption{All of the blocks in HPChain reach consensus: the \texttt{HPChain-total} line is completely covered by the \texttt{HPChain-bc} line.
    }
    \label{fig:block-consensus}
        \figspace
\end{figure}

To further study the reliability of HPChain, we investigate how many blocks reach consensus in all the systems on 16 nodes. The reason for choosing 16 nodes is to keep the result comparable to ~\cite{blockbench_sigmod17}. Figure~\ref{fig:block-consensus} reports the number of blocks that reach the consensus and are appended to the blockchain. Both in Ethereum and Parity, some blocks are unable to reach consensus due to double spending or selfish mining attacks, and the difference increases as time passes. Though Hyperledger is not vulnerable to those attacks because of not allowing any fork and $100\%$ blocks tend to reach consensus, it is significantly slower than HPChain. In HPChain, almost $100\times$ more blocks are generated compared to Hyperledger, and $100\%$ blocks reach consensus. This is because, HPChain implementation relies on a modified version of consensus protocol (i.e., HyBE) that is extended from PoW (i.e., low difficulty in computation) and PBFT (i.e., parallel message passing with MPI) to ensure block validity in a parallel manner.

\begin{figure}[!t]
    \centering
    \includegraphics[width=75mm]{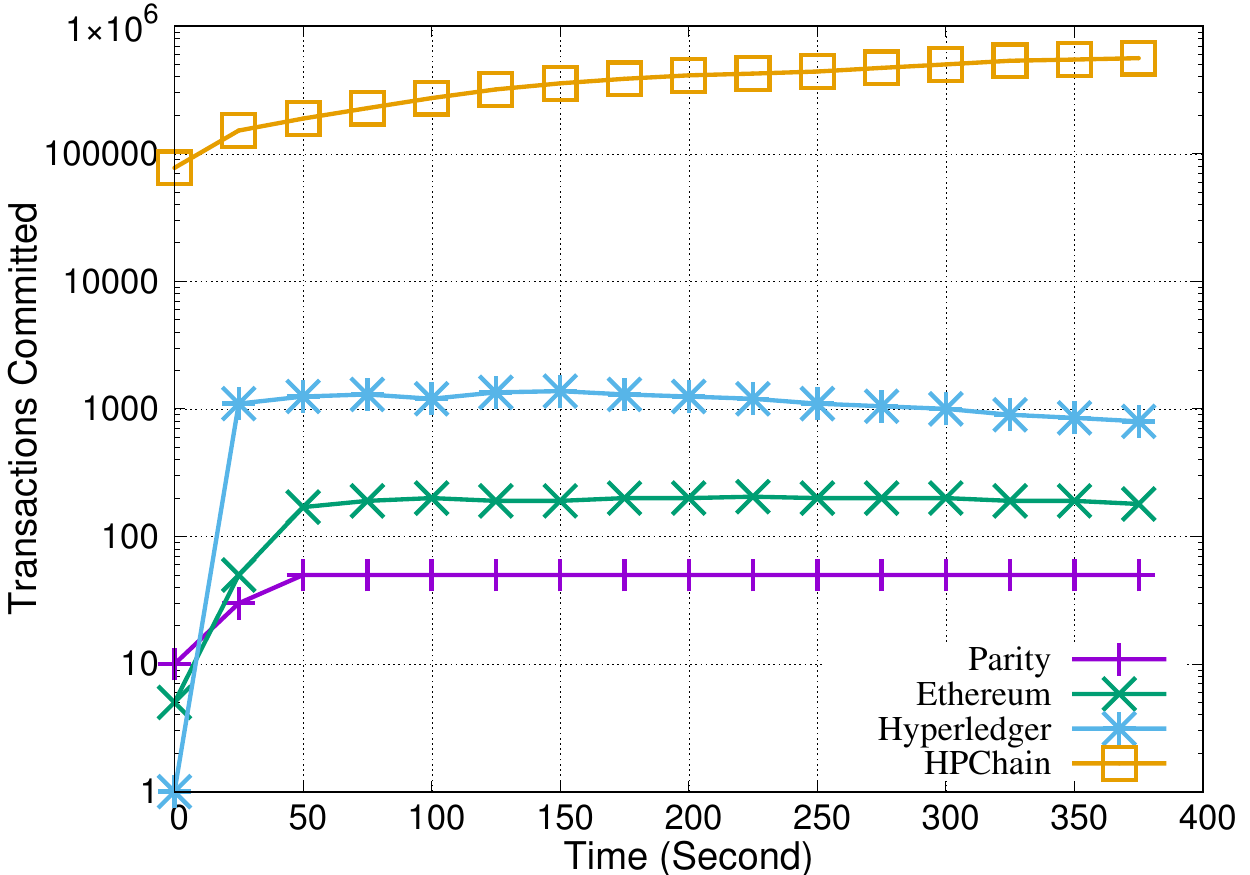}
    \caption{Transactions throughput is not degraded by the node failure in HPChain.
    }
    \label{fig:txn-committed}
        \figspace
\end{figure}

To investigate the fault tolerance, we check the transaction throughput in HPChain over 400 seconds on 20 nodes and compare with the result of other systems~\cite{blockbench_sigmod17}. During the experiment, we intentionally switch off four nodes to check the performance of HPChain. As shown in Figure~\ref{fig:txn-committed}, HPChain exhibits significant tolerance without affecting throughput over the execution time. In both Parity and Ethereum, the performance remains unaffected; therefore, failed servers do not affect the throughput. However, in Hyperledger, throughput starts to drop considerably after 250 seconds. This is because, with 16 nodes, Hyperledger starts to generate blocks at a slower rate as PBFT can not tolerate more than four failures. However, it is noticeable that switching of 4 nodes in HPChain does not affect the throughput increment, and instead of getting low or remaining constant, the throughput increases moderately with time because HPChain utilizes parallel message passing to commit transactions and remote storage to continue validation during the node failures. 

\begin{figure}[!t]
    \centering

\includegraphics[width=75mm]{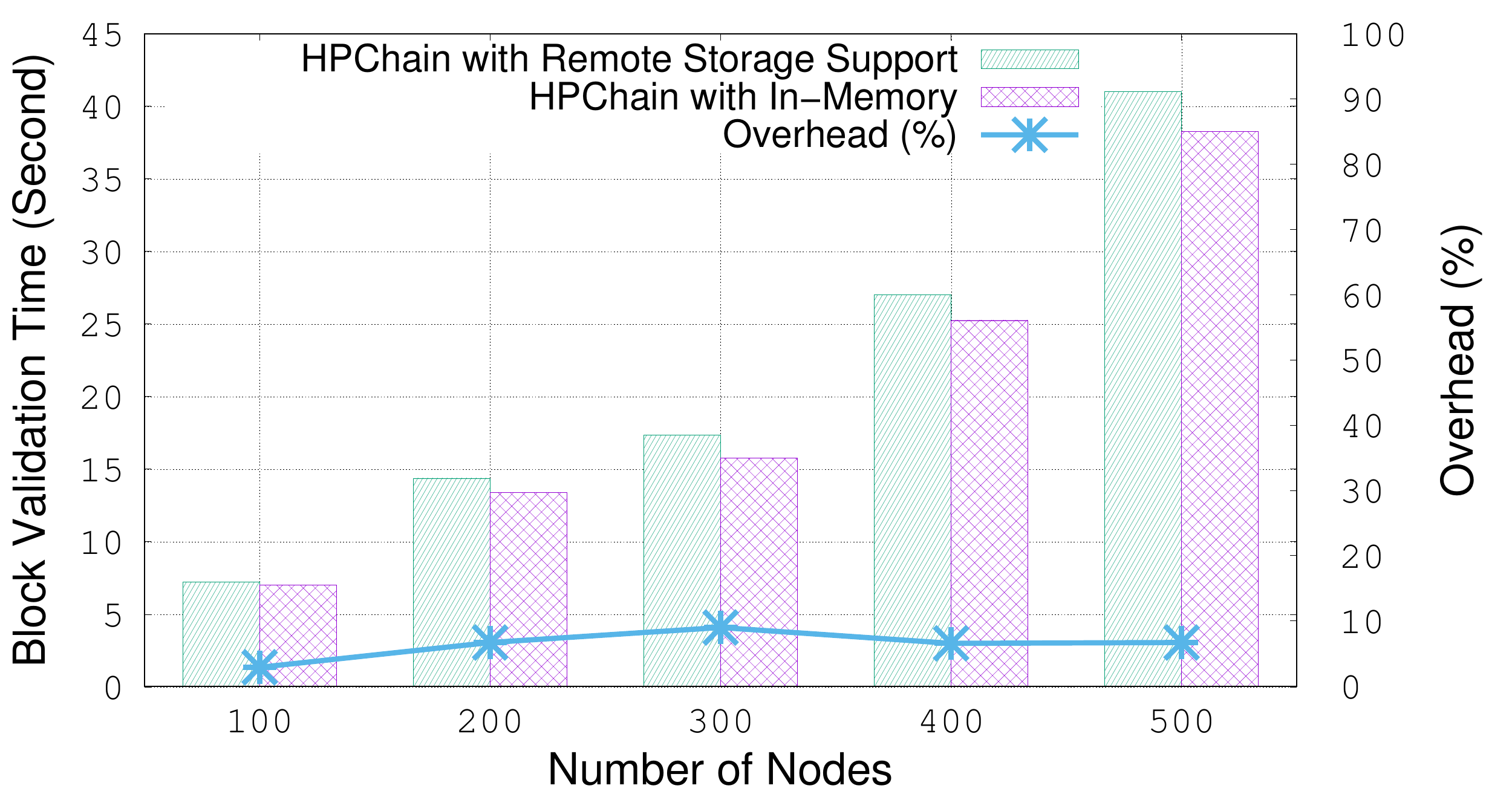}

    \caption{The remote storage introduces insignificant overhead to the performance of HPChain during node failure.}
    \label{fig:ledger-ft}
        \figspace
\end{figure}

To further study the fault tolerance, we check the block validation time by HPChain both with remote storage and in-memory support at different scales by switching off a random node in the middle of a block validation process and compare the overhead. The validation time for HPChain with remote ledger presented here is the total time needed to validate the block with at-least 51\% compute nodes and with the remote ledger. We also measure the time required to complete the validation process with only in-memory support. As shown in Figure~\ref{fig:ledger-ft}, even with failures of the compute node, the block validation process continues successfully. Besides, HPChain with remote ledger incurs negligible overhead compared to the HPChain in-memory support even at a large scale (i.e., roughly 5\% at 500 nodes cluster).

\begin{figure}[!t]
    \centering
    \includegraphics[width=75mm]{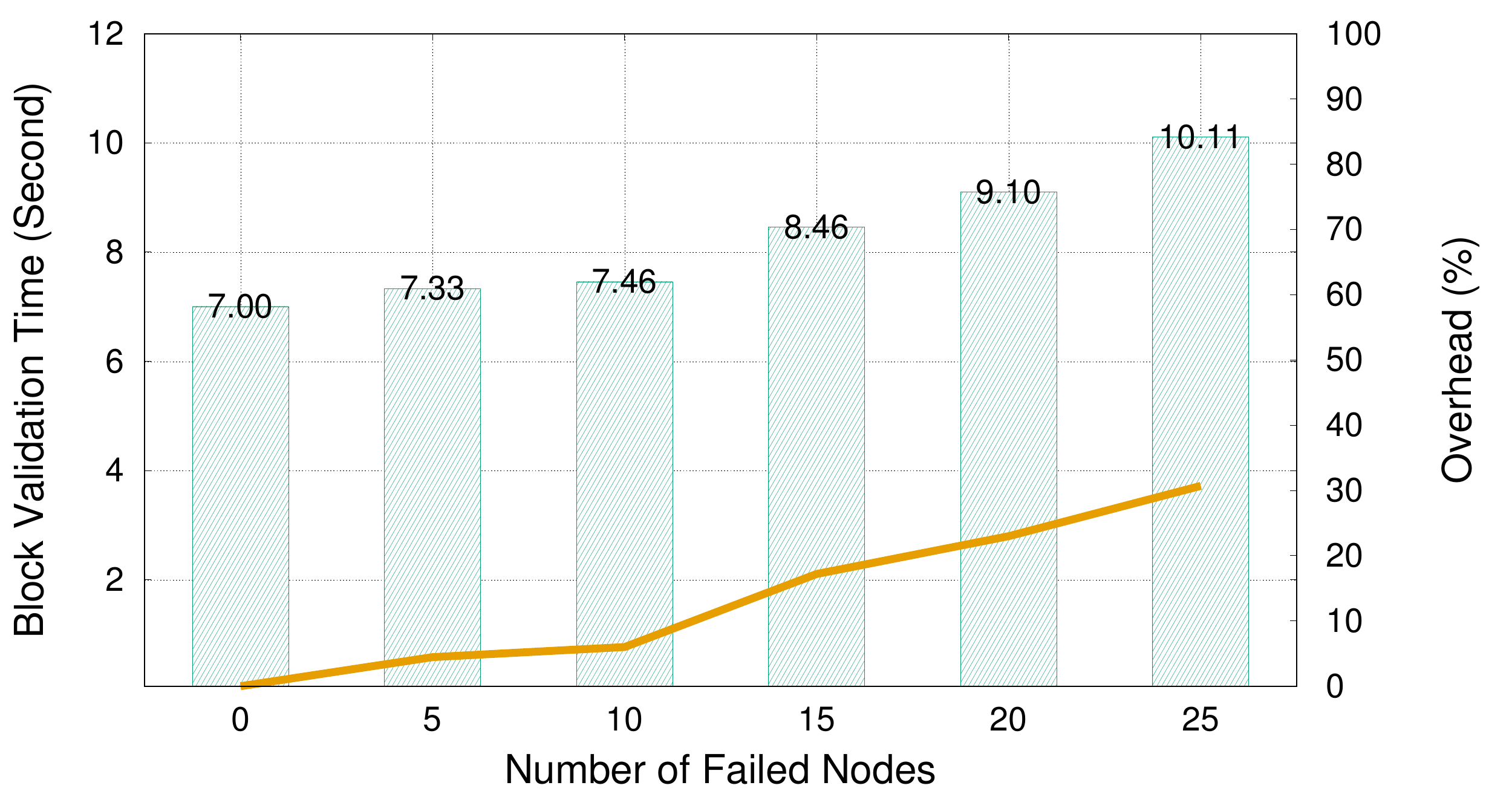}
    
    \caption{On a 100-node cluster, the performance overhead introduced by the validation at the remote storage.}
    \label{fig:ledger-faulttolerance-with-failed-nodes}
        \figspace
\end{figure}

To analyze the fault tolerance mechanism in HPChain under several nodes failure, we further continue our experiment with 100 nodes cluster, where we intentionally switch off a different number of nodes while validating a block. The goal of this experiment is to see how much is the overhead if we leverage the remote storage for each node crash.  That is, for each node failure, the validation process is redirected to the remote storage for successful continuation.  Figure~\ref{fig:ledger-faulttolerance-with-failed-nodes} shows that even with several failed nodes, the fault monitor in HPChain helps to continue the block validation process successfully with negligible overhead (i.e., nearly 5\%) on up to 10 nodes failure. The overhead increases with the number of failed nodes because the remote storage is accessed for each node failure in this specific experiment. However, in the actual fault tolerance mechanism implemented in HPChain, the remote ledger is accessed only once when 51\% nodes fail. Therefore, the scaling of nodes does not have any significant impact on the overhead of accessing the remote storage.

\section{Conclusion and Future Work}
\label{sec:conclusion}
This paper proposes a new blockchain system architecture with a parallel mechanism based on MPI, where ledgers are maintained in parallel mostly in memory.
Besides, we propose a set of consensus protocols crafted for the new parallel architecture. 
The new architecture and consensus protocols,
under the framework coined as BAASH,
collectively enable an efficient parallel blockchain-like ledger service for trustworthy data management on HPC systems. 
The validity of the proposed consensus is experimentally verified. A light-weight system prototype is implemented and evaluated with more than two million transactions on a 500-core HPC cluster.
The evaluation shows BAASH delivers $6\times$, $12\times$, and $75\times$ higher throughput than mainstream blockchains Hyperldeger, Ethereum, and Parity, respectively.

At the writing of this paper, we are working with our collaborators to deploy BAASH to one of the top supercomputers from the Top-500 list~\cite{top500}. 
We also plan to evaluate BAASH using a recently published emulator for blockchains called BlockLite~\cite{xwang_cloud19},
which can accurately emulate blockchain protocols on up tp to 40,000 nodes.
In addition to further pushing the scalability toward thousands and even tens of thousands of nodes,
an important topic, 
under which this project is currently funded by a U.S. federal agency, 
is to take into account the energy consumption into our current design space and upgrade BAASH to guarantee strong energy efficiency, 
both theoretically and experimentally.

\bibliographystyle{plain}
\bibliography{ref_new}

\begin{thebibliography}{10}

\bibitem{ornl_bc18}
{Advancing the Science and Impact of Blockchain Technology at Oak Ridge
  National Laboratory}.
\newblock \url{https://info.ornl.gov/sites/publications/Files/Pub118487.pdf},
  Accessed 2019.

\bibitem{aalmamun_sc19}
Abdullah Al-Mamun, Tonglin Li, Mohammad Sadoghi, Linhua Jiang, Haoting Shen,
  and Dongfang Zhao.
\newblock Poster: An mpi-based blockchain framework for data fidelity in
  high-performance computing systems.
\newblock In {\em International Conference on High Performance Computing,
  Networking, Storage and Analysis (SC)}, 2019.

\bibitem{aalmamun_bigdata18}
Abdullah Al-Mamun, Tonglin Li, Mohammad Sadoghi, and Dongfang Zhao.
\newblock In-memory blockchain: Toward efficient and trustworthy data
  provenance for hpc systems.
\newblock In {\em IEEE International Conference on Big Data (BigData)}, 2018.

\bibitem{aniello2017}
L.~Aniello, R.~Baldoni, E.~Gaetani, F.~Lombardi, A.~Margheri, and V.~Sassone.
\newblock A prototype evaluation of a tamper-resistant high performance
  blockchain-based transaction log for a distributed database.
\newblock In {\em 13th European Dependable Computing Conference (EDCC)}, 2017.

\bibitem{bigchaindb}
{BigchainDB}.
\newblock \url{https://github.com/bigchaindb/bigchaindb}, Accessed 2018.

\bibitem{bitcoin}
{Bitcoin}.
\newblock \url{ https://bitcoin.org/bitcoin.pdf}, Accessed 2018.

\bibitem{bitcointy}
{Bitcoin Scale}.
\newblock \url{
  https://data.bitcoinity.org/bitcoin/block_time/5y?f=m10&r=month&t=l},
  Accessed 2019.

\bibitem{mpi-consensus-buntinas2012}
Darius Buntinas.
\newblock Scalable distributed consensus to support mpi fault tolerance.
\newblock In {\em 2012 IEEE 26th International Parallel and Distributed
  Processing Symposium}, pages 1240--1249. IEEE, 2012.

\bibitem{mpi-hive-chao2015}
Lu~Chao, Chundian Li, Fan Liang, Xiaoyi Lu, and Zhiwei Xu.
\newblock Accelerating apache hive with mpi for data warehouse systems.
\newblock In {\em 2015 IEEE 35th International Conference on Distributed
  Computing Systems}, pages 664--673. IEEE, 2015.

\bibitem{mpi-Chunduri-2018}
Sudheer Chunduri, Scott Parker, Pavan Balaji, Kevin Harms, and Kalyan Kumaran.
\newblock Characterization of mpi usage on a production supercomputer.
\newblock In {\em Proceedings of the International Conference for High
  Performance Computing, Networking, Storage, and Analysis}, SC '18, pages
  30:1--30:15. IEEE Press, 2018.

\bibitem{blockbench_sigmod17}
Tien Tuan~Anh Dinh, Ji~Wang, Gang Chen, Rui Liu, Beng~Chin Ooi, and Kian-Lee
  Tan.
\newblock Blockbench: A framework for analyzing private blockchains.
\newblock In {\em ACM International Conference on Management of Data (SIGMOD)},
  2017.

\bibitem{doe_sbir_2017}
{DOE SBIR}.
\newblock \url{https://www.sbir.gov/sbirsearch/detail/1307745}, Accessed 2017.

\bibitem{ethereum}
{Ethereum}.
\newblock \url{ https://www.ethereum.org/}, Accessed 2018.

\bibitem{crypto-bitcoin-ng-Eyal:2016}
Ittay Eyal, Adem~Efe Gencer, Emin~G\"{u}n Sirer, and Robbert Van~Renesse.
\newblock Bitcoin-ng: A scalable blockchain protocol.
\newblock In {\em Proceedings of the 13th Usenix Conference on Networked
  Systems Design and Implementation}, NSDI'16, pages 45--59, Berkeley, CA, USA,
  2016. USENIX Association.

\bibitem{spade_2012}
Ashish Gehani and Dawood Tariq.
\newblock {SPADE: Support for Provenance Auditing in Distributed Environments}.
\newblock In {\em Proceedings of the 13th International Middleware Conference
  (Middleware)}, 2012.

\bibitem{crypto-algorand-Gilad:2017}
Yossi Gilad, Rotem Hemo, Silvio Micali, Georgios Vlachos, and Nickolai
  Zeldovich.
\newblock Algorand: Scaling byzantine agreements for cryptocurrencies.
\newblock In {\em Proceedings of the 26th Symposium on Operating Systems
  Principles}, SOSP '17, pages 51--68, New York, NY, USA, 2017. ACM.

\bibitem{sbft-gueta2018}
G.~{Golan Gueta}, I.~{Abraham}, S.~{Grossman}, D.~{Malkhi}, B.~{Pinkas},
  M.~{Reiter}, D.~{Seredinschi}, O.~{Tamir}, and A.~{Tomescu}.
\newblock Sbft: A scalable and decentralized trust infrastructure.
\newblock In {\em 2019 49th Annual IEEE/IFIP International Conference on
  Dependable Systems and Networks (DSN)}, 2019.

\bibitem{GROPP1996}
William Gropp, Ewing Lusk, Nathan Doss, and Anthony Skjellum.
\newblock A high-performance, portable implementation of the mpi message
  passing interface standard.
\newblock {\em Parallel Computing}, 22(6):789 -- 828, 1996.

\bibitem{mpi-Guo-2015}
Yanfei Guo, Wesley Bland, Pavan Balaji, and Xiaobo Zhou.
\newblock Fault tolerant mapreduce-mpi for hpc clusters.
\newblock In {\em Proceedings of the International Conference for High
  Performance Computing, Networking, Storage and Analysis}, SC '15, pages
  34:1--34:12, 2015.

\bibitem{msadoghi_ebdt18}
Suyash Gupta and Mohammad Sadoghi.
\newblock {\em Blockchain Transaction Processing}.
\newblock Springer International Publishing, 2018.

\bibitem{hyperledger}
{Hyperledger}.
\newblock \url{ https://www.hyperledger.org/}, Accessed 2018.

\bibitem{inkchain}
{Inkchain}.
\newblock \url{https://github.com/inklabsfoundation/inkchain}, Accessed 2018.

\bibitem{Kiayias_2017}
Aggelos Kiayias, Alexander Russell, Bernardo David, and Roman Oliynykov.
\newblock Ouroboros: A provably secure proof-of-stake blockchain protocol.
\newblock In {\em Advances in Cryptology (CRYPTO)}, 2017.

\bibitem{crypto-kogias2016}
Eleftherios~Kokoris Kogias, Philipp Jovanovic, Nicolas Gailly, Ismail Khoffi,
  Linus Gasser, and Bryan Ford.
\newblock Enhancing bitcoin security and performance with strong consistency
  via collective signing.
\newblock In {\em 25th $\{$USENIX$\}$ Security Symposium ($\{$USENIX$\}$
  Security 16)}, pages 279--296, 2016.

\bibitem{bizcoin-kogias2016enhancing}
Eleftherios~Kokoris Kogias, Philipp Jovanovic, Nicolas Gailly, Ismail Khoffi,
  Linus Gasser, and Bryan Ford.
\newblock Enhancing bitcoin security and performance with strong consistency
  via collective signing.
\newblock In {\em 25th $\{$USENIX$\}$ Security Symposium ($\{$USENIX$\}$
  Security 16)}, pages 279--296, 2016.

\bibitem{sharding-omniledger-kokoris2018}
Eleftherios Kokoris-Kogias, Philipp Jovanovic, Linus Gasser, Nicolas Gailly,
  Ewa Syta, and Bryan Ford.
\newblock Omniledger: A secure, scale-out, decentralized ledger via sharding.
\newblock In {\em 2018 IEEE Symposium on Security and Privacy (SP)}, pages
  583--598. IEEE, 2018.

\bibitem{kosba2016}
A.~Kosba, A.~Miller, E.~Shi, Z.~Wen, and C.~Papamanthou.
\newblock Hawk: The blockchain model of cryptography and privacy-preserving
  smart contracts.
\newblock In {\em 2016 IEEE Symposium on Security and Privacy (SP)}, 2016.

\bibitem{slee_icde17}
S.~Lee, S.~Kohler, B.~Ludascher, and B.~Glavic.
\newblock A sql-middleware unifying why and why-not provenance for first-order
  queries.
\newblock In {\em IEEE International Conference on Data Engineering (ICDE)},
  2017.

\bibitem{hpc-li2017proof}
Kejiao Li, Hui Li, Hanxu Hou, Kedan Li, and Yongle Chen.
\newblock Proof of vote: A high-performance consensus protocol based on vote
  mechanism \& consortium blockchain.
\newblock In {\em 2017 IEEE 19th International Conference on High Performance
  Computing and Communications; IEEE 15th International Conference on Smart
  City; IEEE 3rd International Conference on Data Science and Systems
  (HPCC/SmartCity/DSS)}, pages 466--473. IEEE, 2017.

\bibitem{mpi-li2016}
Mingzhe Li, Khaled Hamidouche, Xiaoyi Lu, Hari Subramoni, Jie Zhang, and
  Dhabaleswar~K Panda.
\newblock Designing mpi library with on-demand paging (odp) of infiniband:
  challenges and benefits.
\newblock In {\em SC'16: Proceedings of the International Conference for High
  Performance Computing, Networking, Storage and Analysis}, pages 433--443.
  IEEE, 2016.

\bibitem{xliang_ccgrid17}
X.~Liang, S.~Shetty, D.~Tosh, C.~Kamhoua, K.~Kwiat, and L.~Njilla.
\newblock Provchain: A blockchain-based data provenance architecture in cloud
  environment with enhanced privacy and availability.
\newblock In {\em IEEE/ACM International Symposium on Cluster, Cloud and Grid
  Computing (CCGRID)}, 2017.

\bibitem{mpi-datampi-lu2014}
Xiaoyi Lu, Fan Liang, Bing Wang, Li~Zha, and Zhiwei Xu.
\newblock Datampi: extending mpi to hadoop-like big data computing.
\newblock In {\em 2014 IEEE 28th International Parallel and Distributed
  Processing Symposium}, pages 829--838. IEEE, 2014.

\bibitem{honey-miller2016}
Andrew Miller, Yu~Xia, Kyle Croman, Elaine Shi, and Dawn Song.
\newblock The honey badger of bft protocols.
\newblock In {\em Proceedings of the 2016 ACM SIGSAC Conference on Computer and
  Communications Security}, pages 31--42. ACM, 2016.

\bibitem{mpi}
{MPI}.
\newblock \url{https://www.mpi-forum.org/docs/}, Accessed 2019.

\bibitem{mpi4py}
{MPI4PY}.
\newblock \url{ https://mpi4py.readthedocs.io/en/stable/intro.html}, Accessed
  2019.

\bibitem{nsf_cici_2018}
{NSF CICI}.
\newblock \url{https://www.nsf.gov/pubs/2018/nsf18547/nsf18547.htm}, Accessed
  2018.

\bibitem{SmartGov-olnes2016}
Svein {\O}lnes.
\newblock Beyond bitcoin enabling smart government using blockchain technology.
\newblock In {\em International Conference on Electronic Government}, pages
  253--264. Springer, 2016.

\bibitem{parity}
{Parity}.
\newblock \url{ https://ethcore.io/parity.html/}, Accessed 2018.

\bibitem{hqin_sc19}
Heyang Qin, Syed Zawad, Yanqi Zhou, Lei Yang, Dongfang Zhao, and Feng Yan.
\newblock Swift machine learning model serving scheduling: a region based
  reinforcement learning approach.
\newblock In {\em Proceedings of the International Conference for High
  Performance Computing, Networking, Storage and Analysis (SC)}, 2019.

\bibitem{mpi-ppopp-Saillard2015}
Emmanuelle Saillard, Patrick Carribault, and Denis Barthou.
\newblock Static/dynamic validation of mpi collective communications in
  multi-threaded context.
\newblock In {\em Proceedings of the 20th ACM SIGPLAN Symposium on Principles
  and Practice of Parallel Programming}, PPoPP 2015, pages 279--280. ACM, 2015.

\bibitem{gpfs_2002}
Frank Schmuck and Roger Haskin.
\newblock {GPFS}: A shared-disk file system for large computing clusters.
\newblock In {\em Proceedings of the 1st USENIX Conference on File and Storage
  Technologies (FAST)}, 2002.

\bibitem{sha256}
{SHA-256}.
\newblock \url{https://en.bitcoin.it/wiki/SHA-256}, Accessed 2018.

\bibitem{dzhao_tapp13}
Chen Shou, Dongfang Zhao, Tanu Malik, and Ioan Raicu.
\newblock Towards a provenance-aware distributed filesystem.
\newblock In {\em TaPP Workshop, USENIX Symposium on Networked Systems Design
  and Implementation (NSDI)}, 2013.

\bibitem{jsousa_dsn18}
J.~Sousa, A.~Bessani, and M.~Vukolic.
\newblock A byzantine fault-tolerant ordering service for the hyperledger
  fabric blockchain platform.
\newblock In {\em 48th Annual IEEE/IFIP International Conference on Dependable
  Systems and Networks (DSN)}, 2018.

\bibitem{cori}
{The Cori Supercomputer}.
\newblock \url{http://www.nersc.gov/users/computational-systems/cori}, Accessed
  2018.

\bibitem{top500}
Top500.
\newblock \url{https://www.top500.org/lists/2019/06}, Accessed 2019.

\bibitem{exascale-valduriez2015data}
Patrick Valduriez.
\newblock Data-intensive hpc: opportunities and challenges.
\newblock In {\em BDEC'2015: Big Data and Extreme-scale Computing}, 2015.

\bibitem{Wang_nsdi19}
Jiaping Wang and Hao Wang.
\newblock Monoxide: Scale out blockchain with asynchronized consensus zones.
\newblock In {\em 16th {USENIX} Symposium on Networked Systems Design and
  Implementation ({NSDI} 19)}, Boston, MA, 2019. {USENIX} Association.

\bibitem{xwang_cloud19}
Xinying Wang, Abdullah Al{-}Mamun, Feng Yan, and Dongfang Zhao.
\newblock Toward accurate and efficient emulation of public blockchains in the
  cloud.
\newblock In {\em 12th International Conference on Cloud Computing (CLOUD)},
  pages 67--82, 2019.

\bibitem{ycsb}
{YCSB}.
\newblock \url{ https://github.com/brianfrankcooper/YCSB/wiki/Core-Workloads},
  Accessed 2018.

\bibitem{sharding-rapidchain-zamani2018}
Mahdi Zamani, Mahnush Movahedi, and Mariana Raykova.
\newblock Rapidchain: Scaling blockchain via full sharding.
\newblock In {\em Proceedings of the 2018 ACM SIGSAC Conference on Computer and
  Communications Security}, pages 931--948. ACM, 2018.

\bibitem{kzhang_icdcs18}
Kaiwen Zhang and Hans{-}Arno Jacobsen.
\newblock Towards dependable, scalable, and pervasive distributed ledgers with
  blockchains.
\newblock In {\em 38th {IEEE} International Conference on Distributed Computing
  Systems (ICDCS)}, 2018.

\bibitem{dzhao_cluster13}
Dongfang Zhao, Chen Shou, Tanu Malik, and Ioan Raicu.
\newblock Distributed data provenance for large-scale data-intensive computing.
\newblock In {\em IEEE International Conference on Cluster Computing
  (CLUSTER)}, 2013.

\end{thebibliography}

\clearpage

\end{document}